\documentclass[a4paper,fleqn,usenatbib]{mnras}

\usepackage[T1]{fontenc}
\usepackage{ae,aecompl}

\usepackage{graphicx}	% Including figure files
\usepackage{amsmath}	% Advanced maths commands
\usepackage{amssymb}	% Extra maths symbols
\usepackage{physics}
\usepackage{xcolor}

\title[HMM limits on UTMOST glitches]{Systematic upper limits on the size of missing pulsar glitches in the first UTMOST open data release}

\author[L. Dunn et al.]{
L. Dunn$^{1,2}$\thanks{E-mail: liamd@student.unimelb.edu.au}, A. Melatos$^{1,2}$, S. Suvorova$^{1,2,3}$, W. Moran$^{3}$, R. J. Evans$^{2,3}$, \newauthor S. Os\l owski$^{4}$, M. E. Lower$^{5,6}$, M. Bailes$^{5,7}$, C. Flynn$^{5,7}$, V. Gupta$^{5}$
\\
$^{1}$School of Physics, University of Melbourne, Parkville, VIC 3010, Australia\\
$^{2}$Australian Research Council Centre of Excellence for Gravitational Wave Discovery (OzGrav), University of Melbourne,\\
Parkville, VIC 3010, Australia\\
$^{3}$Department of Electrical and Electronic Engineering, University of Melbourne, Parkville, VIC 3010, Australia\\
$^{4}$Manly Astrophysics, 15/41-42 East Esplanade, Manly 2095, Australia\\
$^{5}$Centre for Astrophysics and Supercomputing, Swinburne University of Technology, PO Box 218, Hawthorn, Victoria 3122, Australia\\
$^{6}$CSIRO Space and Astronomy, Australia Telescope National Facility, Epping NSW 1710, Australia\\
$^{7}$OzGrav: The ARC Centre of Excellence for Gravitational-wave Discovery, Hawthorn VIC 3122, Australia
}

\date{Accepted XXX. Received YYY; in original form ZZZ}

\pubyear{2020}

\begin{document}
\label{firstpage}
\pagerange{\pageref{firstpage}--\pageref{lastpage}}
\maketitle

\begin{abstract}
A systematic, semi-automated search for pulsar glitches in the first UTMOST public data release is presented.
The search is carried out using a hidden Markov model which incorporates both glitches and timing noise into the model of the assumed phase evolution of the pulsar.
Glitches are detected through Bayesian model selection between models with and without glitches present with minimal human intervention.
Nine glitches are detected among seven objects, all of which have been previously reported.
No new glitches were detected.
Injection studies are used to place 90\% frequentist upper limits on the size of undetected glitches in each of the 282 objects searched.
The mean upper limit obtained is $\Delta f^{90\%}/f = 1.9 \times 10^{-8}$, with a range of $4.1 \times 10^{-11} \leq \Delta f^{90\%}/f \leq 2.7 \times 10^{-7}$, assuming step events with no post-glitch recoveries.
It is demonstrated that including glitch recovery has a mild effect, in most cases increasing the upper limit by a factor of $\lesssim 5$ conservatively assuming complete recovery on a timescale of $100\,\mathrm{d}$.
\end{abstract}


%%%%%%%%%%%%%%%%%%%%%%%%%%%%%%%%%%%%%%%%%%%%%%%%%%

%%%%%%%%%%%%%%%%% BODY OF PAPER %%%%%%%%%%%%%%%%%%

\section{Introduction}
The secular electromagnetic spindown of a rotation-powered pulsar is sometimes interrupted by a sudden increase in the spin frequency, known as a glitch.
Glitches are often but not always accompanied by a change in the secular spin-down rate and a quasi-exponential recovery \citep{lynePulsarAstronomy2012}.
The underlying cause of glitches is unknown \citep{haskellModelsPulsarGlitches2015a}.
The standard view of the physical mechanism behind glitches invokes pinning and subsequent unpinning of the vortices of the superfluid neutron component to the lattice of nuclei in the inner crust \citep{andersonPulsarGlitchesRestlessness1975}, but this broad picture is by no means certain.
Long-term statistical analyses have uncovered interesting features of glitch behaviour both in individual objects \citep{espinozaNeutronStarGlitches2014, howittNonparametricEstimationSize2018,carlinAutocorrelationsPulsarGlitch2019, hoReturnBigGlitcher2020} and across the pulsar population \citep{lyneStatisticalStudiesPulsar2000a,melatosAvalancheDynamicsRadio2008,espinozaStudy315Glitches2011, yuDetection107Glitches2013,fuentesGlitchActivityNeutron2017a, melatosSizewaitingtimeCorrelationsPulsar2018}.
Note that we distinguish between glitches, which involve a jump in the spin frequency, and events involving abrupt changes in frequency derivative associated with magnetospheric changes \citep{lyneSwitchedMagnetosphericRegulation2010a}.
Although the latter are glitch-like in some respects and interesting in their own right, they are a distinct class of events and we will not search for them explicitly in this work.

Statistical inferences about the glitch phenomenon rely on the completeness of the catalogues of detected glitches.
However, the traditional method of glitch detection, which involves identifying a glitch signature ``by eye'' in a set of timing residuals \citep{espinozaStudy315Glitches2011, yuDetection107Glitches2013}, makes it difficult to assess completeness systematically.
\citet{espinozaNeutronStarGlitches2014} and \citet{yuDetectionProbabilityNeutron2017} employed automated glitch detection methods to evaluate detectability limits.
The technique presented by \citet{espinozaNeutronStarGlitches2014} has been applied to timing data from the Crab and Vela pulsars \citep{EspinozaAntonopoulou2021}, while the technique presented by \citet{yuDetectionProbabilityNeutron2017} was tested on simulated data but not used to search for glitches in real datasets.
Most recently, \citet{SinghaBasu2021} have developed an real-time glitch detection pipeline which operates with minimal human intervention and incorporated it into the timing programme at the Ooty Radio Telescope.
\citet{SinghaBasu2021} reported initial tests of detectability limits with this pipeline, and as more data become available a clearer picture will emerge of the completeness of the glitch sample reported as part of this programme.

\citet{melatosPulsarGlitchDetection2020} developed a complementary method for pulsar glitch detection which tracks the pulse frequency and frequency derivative with a hidden Markov model (HMM)\footnote{\href{https://github.com/ldunn/glitch_hmm}{https://github.com/ldunn/glitch\_hmm}}.
The HMM selects between models with and without glitches within a Bayesian framework, complementing model selection studies with \textsc{temponest} \citep{lentatiTemponestBayesianApproach2014,shannonCharacterizingRotationalIrregularities2016, parthasarathyTimingYoungRadio2019, lowerUTMOSTPulsarTiming2020}.
The HMM dynamics include secular spin down and stochastic spin wandering (``timing noise''), as well as step changes associated with glitches \citep{melatosPulsarGlitchDetection2020}.
As the HMM detects glitches without human intervention, it is well-suited to analysing a large number of pulsar timing datasets.
Its speed makes it practical to do injection studies to obtain upper limits on the size of undetected glitches, and hence quantify the completeness of the existing and new catalogues.

In this work we search for glitches in the datasets released as part of the UTMOST pulsar timing programme \citep{jankowskiUTMOSTPulsarTiming2019,lowerUTMOSTPulsarTiming2020}.
These datasets were released in March 2020, and contain observations of 300 pulsars taken between January 2014 and August 2019. 
To date, 12 glitches across seven pulsars have been detected as part of this timing programme using traditional methods \citep{lowerUTMOSTPulsarTiming2020}.
In this paper we search for new glitches beyond those discovered to date and set upper limits on the size of undetected glitches in 282 of the 300 pulsars.
The layout of the paper is as follows.
In Section \ref{sec:data} we briefly describe the data.
In Section \ref{sec:method} we describe the HMM and explain how to choose the HMM's control variables and search parameters.
In Section \ref{sec:results} we present the results of the search.
We narrow down the initial list of glitch detections through a veto procedure, and follow up the survivors with a refined HMM analysis to determine the basic glitch parameters: glitch epoch and glitch size.
Finally in Section \ref{sec:ul} we present systematic upper limits on the size of undetected glitches in the UTMOST data release.

\section{Data}
\label{sec:data}
The UTMOST pulsar timing programme is an ongoing campaign conducted at the Molonglo Observatory Synthesis Telescope, a pair of $778\,\mathrm{m}$ long east-west cylindrical paraboloid reflectors located near Canberra, Australia \citep{bailesUTMOSTHybridDigital2017}. 
We searched a subset of the data from the first public release\footnote{\href{https://github.com/Molonglo/TimingDataRelease1/}{https://github.com/Molonglo/TimingDataRelease1/}} \citep{lowerUTMOSTPulsarTiming2020}.
The data consist of times of arrival (ToAs) for 300 pulsars, mostly recorded between October 2015 and August 2019, as well as best-fit timing models.
We search for glitches only in the 283 pulsars that are not in binary systems, due to difficulties in extracting ToAs from \textsc{tempo2} which are referenced to the reference frame of the pulsar, rather than the solar system barycentre.

The volume and density of available timing data vary significantly between pulsars.
Table \ref{tab:data_stats} summarises the variation in observing timespan, cadence, and number of ToAs available across the population of UTMOST pulsars.
The observing timespan is the time between the first available ToA and the last available ToA for each object, and the cadence is the mean time elapsed between consecutive ToAs.
A full description of the observation, data reduction and timing analysis procedures is given by \citet{jankowskiUTMOSTPulsarTiming2019}.

\begin{table}
    \centering
    \caption{Observational statistics for the first UTMOST public data release. All quantities are calculated on a per-pulsar basis, and we take the minimum, mean and maximum over the complete set of pulsars in the data release.}
    \label{tab:data_stats}
    \begin{tabular}{lccc}
        \hline
        & Minimum& Mean& Maximum \\\hline
        Observing timespan ($\mathrm{d}$) & 268 & 1054 & 2024 \\
        Cadence ($\mathrm{d}$) & 1.4 & 16 & 49 \\
        Number of ToAs & 25 & 107 & 1458\\\hline
    \end{tabular}
\end{table}

\section{Hidden Markov model}
\label{sec:method}
The implementation of a HMM used to search for glitches is described in detail by \citet{melatosPulsarGlitchDetection2020}.
Here we provide a brief description of the most pertinent aspects.
Section \ref{subsec:trans_emis} discusses the probabilities which determine the dynamics of the HMM, and its connection to the observed data.
Section \ref{subsec:model_selec} introduces the Bayesian model selection procedure which is used to select between models with and without glitches present.
In Section \ref{subsec:doi} we discuss the boundaries and discretisation of the state space of the HMM.
The choices for the various parameters are summarised in Table \ref{tab:hmm_params_initial}.
These are the choices used in the initial searches for previously unknown glitches (Section \ref{sec:results}) and in setting upper limits on the sizes of undetected glitches (Section \ref{sec:ul}).
Analyses aimed at estimating parameters of detected glitches (Sections \ref{subsec:results_J0835}--\ref{subsec:results_J1740}) may require different parameter choices on a case-by-case basis, usually in the allowed range of frequencies, which may need to be extended by more than an order of magnitude to accomodate the glitch.

\begin{table*}
    \centering
    \caption{Domain of interest of physical parameters and HMM control parameters.}
    \begin{tabular}{llll}
        \hline
        Parameter & Symbol & Units & Value \\\hline
        Timing model reference epoch & $T_0$ & MJD & From UTMOST\\
        Secular frequency & $f_\text{LS}$ & Hz & From UTMOST\\
        Secular frequency derivative & $\dot{f}_\text{LS}$ & $\mathrm{Hz}\,\mathrm{s}^{-1}$  & From UTMOST\\
        Frequency deviation & $[f_-, f_+]$ & Hz & $[-3, 3] \times 10^{-7}$ \\
        Frequency derivative deviation & $[\dot{f}_-, \dot{f}_+]$ & $\mathrm{Hz}\,\mathrm{s}^{-1}$ & $[\max(0.1\dot{f}_\text{LS}, -10^{-14}), \min(-0.1\dot{f}_\text{LS}, 10^{-14})]$\\
        Frequency bin size & $\eta_f$ & Hz & $4.0 \times 10^{-10}$ \\
        Frequency derivative bin size & $\eta_{\dot{f}}$ & $\mathrm{Hz}\,\mathrm{s}^{-1}$ & $(\dot{f}_+ - \dot{f}_-)/11$  \\
        Timing noise strength & $\sigma$ & $\mathrm{Hz}\,\mathrm{s}^{-3/2}$ & $\max(10^{-21},\eta_{\dot{f}}\langle x_n\rangle^{-1/2})$ \\
        %Von Mises concentration & $\kappa$ & None & See equation (\ref{eqn:kappa}) &\\
        ToA uncertainty & $\sigma_\mathrm{ToA}$ & s & From UTMOST \\
        Bayes factor threshold & $K_\mathrm{th}$ & None & $10^{1/2}$ \\
        \hline
    \end{tabular}
    \label{tab:hmm_params_initial}
\end{table*}

\subsection{Transition and emission probabilities}
\label{subsec:trans_emis}
A HMM is an automaton which transitions stochastically between a set of hidden states at discrete times $t_1, \ldots, t_{N_T}$, which are spaced unequally in general.
The states are hidden in the sense that they cannot be observed directly; the state of the system must be inferred from observations of auxiliary variables related probabilistically to the hidden states rather than the hidden states themselves.
The probability of jumping from state $q_i$ at time $t_n$ to state $q_j$ at $t_{n+1}$, which is called the transition probability $A_{q_j q_i}(t_n)$, depends only on the state at $t_n$ by the Markov property.
The probability that the system occupies the state $q_i$ at time $t_n$, given an observational datum $o(t_n)$ collected at the same time, is called the emission probability $L_{o(t_n) q_i}$.
The prior, $\Pi_{q_i}$, is the probability that the system is initialized in the state $q_i$.
Together, $A_{q_j q_i}(t_n)$, $L_{o(t_n) q_i}$, and $\Pi_{q_i}$ define a HMM uniquely.

To apply a HMM to pulsar timing, we identify the hidden states $q_i$ with a discrete grid of $(f, \dot{f})$ pairs, which encode the insantaneous spin frequency and its first time derivative.
The hidden $(f, \dot{f})$ states are combined with fixed secular values $f_\text{LS}$ and $\dot{f}_\text{LS}$ measured at a reference epoch $T_0$ to give the instantaneous spin frequency $[f_\text{LS} +\dot{f}_\text{LS}(t_n - T_0)] + f$ and time derivative $\dot{f}_\text{LS} + \dot{f}$ respectively at time $t_n$.
In this work we measure $f_\text{LS}$ and $\dot{f}_\text{LS}$ using \textsc{tempo2} \citep{hobbsTEMPO2NewPulsartiming2006}.
We note that while the HMM requires a measurement of $f_\text{LS}$ and $\dot{f}_\text{LS}$, these values may be derived only from a small subsection of the data in cases where a phase-connected solution spanning the whole dataset is not available.
Although phase-connected solutions are available for all of the UTMOST pulsars searched in this work, the HMM does not incorporate the provided pulse numbering information.
The states can be enlarged to include the second time derivative $\ddot{f}$ (and higher derivatives), but systematic validation tests with real and synthetic data indicate that state enlargement is unnecessary for the application in this paper \citep{melatosPulsarGlitchDetection2020}.
The observational datum at each time $t_n$ is the ToA difference $o(t_n) = t_n - t_{n-1}$ [for ease of notation we write $x_n \equiv o(t_n)$ in the remainder of this paper], which is related probabilistically to the hidden states.%; for example, $f(t_n)$ must be such that an integer number of pulses arrive in the interval $[t_{n-1}, t_n]$, and its inferred value is not unique.
The prior is deliberately chosen to be flat, i.e. $\Pi_{q_i}$ is constant within a restricted parameter domain (see Section \ref{subsec:doi}), as $q_i(t_1)$ is unknown and astrophysically irrelevant.
Other structures for the HMM are possible, of course, and the reader interested in pulsar timing is encouraged to experiment with them \citep{rabinerTutorialHiddenMarkov1989}.

In the HMM framework, the probability of observing a particular ToA gap $x_n$ if the hidden state of the pulsar is $(f, \dot{f})$ depends on the accumulated rotational phase over the gap, $\Phi(t_n; f, f_\text{LS}, \dot{f}, \dot{f}_\text{LS}, T_0)$ [denoted $\Phi(t_n; \ldots)$ for brevity].
Note that $\Phi(t_n; \ldots)$ satisfies $0 \leq \Phi(t_n; \ldots) \leq 1$ over one period, i.e. it is in units of cycles, not radians.
The full expression for $\Phi(t_n; \ldots)$ is \begin{equation} \Phi(t_n;\ldots) = [f + f_\text{LS} + \dot{f}_\text{LS} (t_n - T_0)]x_n - \frac{1}{2}(\dot{f}_\text{LS} + \dot{f})x_n^2. \label{eqn:hmm_accum_phase}\end{equation}% where $t_n$ is the epoch corresponding to the end of the ToA gap (\emph{not} the length of the gap).
The minus sign in the second term arises because we are employing a \emph{backwards} Taylor expansion.
Equation (\ref{eqn:hmm_accum_phase}) can be generalised to include a secular second frequency derivative $\ddot{f}_0$, if required \citep{melatosPulsarGlitchDetection2020}.

If $\Phi(t_n; \ldots)$ is close to an integer, the probability of observing the ToA gap $x_n$ is high.
This is quantified via a von Mises distribution, in which the probability of observing $z$ given a hidden state $(f, \dot{f})$ is given by \begin{equation} L_{x_n q_i} = \frac{\exp\{\kappa\cos[2\pi\Phi(t_n; \ldots)]\}}{2\pi I_0(\kappa)}, \end{equation}
where $I_0(x)$ is the zeroth modified Bessel function of the first kind, and $\kappa$ is a parameter known as the concentration.
Roughly speaking $\kappa$ can be thought of as the reciprocal of the variance of $2\pi\Phi(t_n; \ldots)$.
There are two main contributions to variance in $\Phi(t_n; \ldots)$: measurement uncertainty in the ToAs, and the spacing in the discretized $f$-$\dot{f}$ grid.
If the uncertainties in the ToAs at the beginning and end of the gap are $\sigma_{\text{ToA},1}$ and $\sigma_{\text{ToA},2}$ respectively then the contribution to the phase variance is $f_\text{LS}^2\left(\sigma_{\text{ToA},1}^2 + \sigma_{\text{ToA},2}^2\right)$.
Given spacings in $f$ and $\dot{f}$ of $\eta_f$ and $\eta_{\dot{f}}$, the respective contributions to the phase variance are $(\eta_f x_n)^2$ and $(\eta_{\dot{f}}x_n^2/2)^2$.
Combining these contributions in quadrature, we arrive at\footnote{The factor $(2\pi)^{-2}$ in (\ref{eqn:kappa_defn}) was ommitted accidentally by \citet{melatosPulsarGlitchDetection2020} in equations (8) and (C3) of the latter reference.} \begin{align} \kappa = (2\pi)^{-2}[&f_\text{LS}^2\left(\sigma_{\text{ToA},1}^2 + \sigma_{\text{ToA},2}^2\right)\nonumber \\&+ (\eta_f x_n)^2 + (\eta_{\dot{f}}x_n^2/2)^2 ]^{-1}. \label{eqn:kappa_defn} \end{align}

%This method of glitch detection is based on an approach to pulsar timing in which the evolution of the pulsar frequency and frequency derivative is modelled with an HMM
%For each gap between consecutive ToAs, a model is created in which the pulsar evolves normally (i.e. evolves due to timing noise and any secular spindown terms which are included) except at the beginning of the specified gap, where a glitch occurs.
During each gap between consecutive ToAs, we assume that the pulsar's hidden state evolves stochastically due to timing noise in the absence of a glitch.
The form of the timing noise determines the tarnsition probability $A_{q_j q_i}$ and is unknown a priori for any individual pulsar.
One reasonable model, introduced by \citet{melatosPulsarGlitchDetection2020} and tested satisfactorily on real data \citep{melatosPulsarGlitchDetection2020,LowerJohnston2021} but certainly not unique, assumes that the timing noise is driven by a white-noise torque derivative, \begin{equation} \frac{\mathrm{d}^2f}{\mathrm{d}t^2} = \xi(t), \label{eqn:tn_defn} \end{equation} where $\xi(t)$ is a Langevin term satisfying \begin{equation}\langle\xi(t)\xi(t')\rangle = \sigma^2\delta(t-t') \label{eqn:sigma_defn}\end{equation} and $\sigma$ is the parameter which controls the strength of the timing noise.
Equations (\ref{eqn:tn_defn}) and (\ref{eqn:sigma_defn}) lead to a robust HMM with easy-to-specify transition probabilities $A_{q_j q_i}(t_n)$.
An explicit expression for $A_{q_j q_i}(t_n)$ is given in equations 10--13 and B7--B11 of \citet{melatosPulsarGlitchDetection2020}.
The choice of (\ref{eqn:tn_defn}) and (\ref{eqn:sigma_defn}) is pragmatic.
Pulsars are not expected to obey (\ref{eqn:tn_defn}) and (\ref{eqn:sigma_defn}) exactly for many reasons.
For example, (\ref{eqn:tn_defn}) and (\ref{eqn:sigma_defn}) produce Brownian motion in the torque, whereas there is observational evidence that some pulsars exhibit Brownian motion in the frequency, i.e. $\mathrm{d}f/\mathrm{d}t = \xi(t)$ \citep{cordesPulsarTimingIII1980,cordesJPLPulsarTiming1985,parthasarathyTimingYoungRadio2019}.

The choice of $\sigma$ is also pragmatic and certainly not unique.
A detailed study of how to optimize $\sigma$ on a per-pulsar basis within the HMM framework is beyond the scope of this work; a first pass at some rules of thumb is found in \citet{melatosPulsarGlitchDetection2020}.
Here we follow \citet{melatosPulsarGlitchDetection2020} in adopting a simple prescription which is based on the fact that the HMM only tracks $\dot{f}$ to a certain resolution, $\eta_{\dot{f}}$ (see Section \ref{subsec:doi}).
We demand that the necessary ``correction'' due to binning in the evolution of $f$ across a ToA gap of length $x_n$ is smaller than the dispersion in $f$ caused by the random walk described in (\ref{eqn:tn_defn}) and (\ref{eqn:sigma_defn}), which implies \begin{equation} \sigma = \eta_{\dot{f}} \langle x_n \rangle^{-1/2}, \label{eqn:sigma_rot}\end{equation} where $\langle x_n \rangle$ is the average length of ToA gaps per pulsar.
This ensures that the discretisation of $\dot{f}$ does not lead to false alarms, although in some cases it may degrade the performance of the glitch detector by inflating needlessly the strength of timing noise included in the model.
Any degradation in sensitivity due to this effect is reflected in the upper limits calculated in Section \ref{sec:ul}.
Because the effect of discretisation of $\dot{f}$ is absorbed into the timing noise in the model, we do not expect a significant effect on inferences made on the value of $f$ (e.g the pointwise most likely sequences $\hat{f}(t_n)$ discussed in Section \ref{sec:results}).
The mathematical form of (\ref{eqn:sigma_rot}) is justified in Section 6.1 of \citet{melatosPulsarGlitchDetection2020}.
We additionally follow \citet{melatosPulsarGlitchDetection2020} and impose a lower bound $\sigma \geq 10^{-21}\,\mathrm{Hz}\,\mathrm{s}^{-3/2}$ to avoid numerical underflow.
The prescription described here and listed in Table \ref{tab:hmm_params_initial} has been successfully tested on synthetic and real data which explicitly does not conform to the timing noise model of equations (\ref{eqn:tn_defn}) and (\ref{eqn:sigma_defn}) \citep{melatosPulsarGlitchDetection2020, LowerJohnston2021}.

If the model includes a glitch during a given ToA gap, then the evolution of the hidden state must be modified accordingly.
We adopt the unrestrictive prescription of \citet{melatosPulsarGlitchDetection2020}: a glitch consists of a positive frequency increment and a possible change in frequency derivative which is allowed to be positive or negative.
Explicitly, if the pulsar is in the hidden state $(f, \dot{f})$ at the beginning of a ToA gap of length $x_n$, then it is allowed to transition with equal probability to any state $(f', \dot{f}')$ as long as one has $f' > f + \dot{f}x_n$.
No restriction is placed on the value of $\dot{f}'$ (within the boundaries specified in Section \ref{subsec:doi}). 

\subsection{Model selection}
\label{subsec:model_selec}
A HMM is a Bayesian inference tool.
It works with the fundamental quantity \begin{align} \Pr(Q_{1:N_T} \mid O_{1:N_T}) = &\Pi_{q(t_1)}L_{o(t_1)q(t_1)}\nonumber\\&\times\prod_{n=2}^{N_T}A_{q(t_n)q(t_{n-1})}L_{o(t_n)q(t_n)}, \end{align} which is the probability that the system occupies the hidden state sequence $Q_{1:N_T} = \left\{q(t_1), \ldots, q(t_{N_T})\right\}$ given the observation sequence $O_{1:N_T} = \left\{o(t_1), \ldots, o(t_{N_T})\right\}$ and a model $M = \left\{A_{q_j q_i}, L_{o(t_n) q_i}, \Pi_{q_i}\right\}$.
The model with no glitch is denoted $M_0$, and the model with a glitch during the $k$th ToA gap is denoted $M_1(k$).
For a given model $M$ and timing data $D = O_{1:N_T}$ we calculate the model evidence $\Pr(D\mid M)$ using the HMM forward algorithm \citep{rabinerTutorialHiddenMarkov1989}.
We can then calculate the ratios \begin{equation} K_1(k) = \frac{\Pr[D \mid M_1(k)]}{\Pr(D \mid M_0)}, \end{equation}
for $1 \leq k \leq N_T$, which are Bayes factors, indicating support for each of the $N_T$ glitch-containing models over the no-glitch model.
According to Bayes's theorem, the ratio of posterior probabilities of the two models includes an extra factor containing the prior probabilities, \begin{equation}\frac{\Pr[M_1(k) \mid D]}{\Pr(M_0 \mid D)} = \frac{\Pr[D \mid M_1(k)]}{\Pr(D \mid M_0)}\frac{\Pr(M_0)}{\Pr[M_1(k)]}. \end{equation} 
Here we make the simplifying assumption $\Pr[M_1(k)] = \Pr(M_0)$ for all $k$, and so the Bayes factor $K_1(k)$ and the ratio of posterior probabilities coincide.
If $\max_{1 \leq k \leq N_T} K_1(k)$ exceeds a pre-defined threshold $K_{\rm th}$, we say that we have a glitch candidate.
We note briefly that the permissive glitch model used may also accomodate possible abrupt changes in spin-down state \citep{lyneSwitchedMagnetosphericRegulation2010a}, and thus model selection may produce glitch candidates associated with these events as well as more typical glitch events.

To account for possible multiple glitches in a dataset, we adopt the greedy hierarchical approach described in Section 4.2 of \citet{melatosPulsarGlitchDetection2020}.
If a glitch candidate is detected when comparing the models $M_1(k)$ to $M_0$, we set $k_1^* = \mathrm{arg max}_k K_1(k)$, and then calculate the ratios \begin{equation} K_2(k_2) = \frac{\Pr[D \mid M_2(k_1^*, k_2)]}{\Pr[D \mid M_1(k_1^*)]}, \end{equation}
where $M_2(k_1^*, k_2)$ is the model containing two glitches at the $k_1^*$ and $k_2$th ToA gaps.
If a second glitch candidate is detected, i.e. $\max_{1 \leq k_2 \leq N_T} K_2(k_2) > K_\text{th}$, we repeat the procedure, now comparing $M_3(k_1^*, k_2^*, k_3)$ against $M_2(k_1^*, k_2^*)$.
The procedure repeats until none of the Bayes factors exceeds $K_\text{th}$.

In order to follow up each candidate, we calculate the posterior distribution of frequency and frequency derivative states during each ToA gap using the HMM forward-backward algorithm \citep{rabinerTutorialHiddenMarkov1989}.
The logic behind this step is discussed in detail in Section 4.3 and Appendix A of \citet{melatosPulsarGlitchDetection2020}.
This posterior distribution can then be used to infer the sequence of most likely frequency states, and hence the most likely size of the frequency jump due to the glitch.

The Bayes factor threshold determines when we have a glitch candidate to be followed up with further analysis.
Here we adopt a fixed threshold of $10^{1/2}$, motivated by the synthetic data tests presented in Section 6 of \citet{melatosPulsarGlitchDetection2020}.
$K_\text{th} = 10^{1/2}$ gives a false alarm probability of roughly 1\%, provided that the timing noise is not much stronger than what is included in the HMM.

\subsection{Domain of interest}
\label{subsec:doi}
The domain of interest (DOI) refers to the set of hidden states which are included in the HMM.
As in \citet{melatosPulsarGlitchDetection2020}, we consider only DOIs which form a grid in a ``reasonable'' $f$--$\dot{f}$ region, which is restricted to avoid wasteful computation; the state sequence $Q_{1:N_T}$ cannot wander unreasonably far from the \textsc{tempo2} fit $f_\text{LS}$ and $\dot{f}_\text{LS}$, because timing noise and glitches represent modest perturbations on the secular trend.
The choices to be made are then the boundaries of the region, and the spacing between points in the grid.
The typical DOI parameters used in this work are summarised in Table \ref{tab:hmm_params_initial}.

When searching for unknown glitches and setting upper limits on the size of undetected glitches, the boundary of the $f$ region is chosen to be the same for all pulsars in this study: the region covered is $-3 \leq f/\left(10^{-7}\,\mathrm{Hz}\right) \leq 3$.
This range generously brackets the typical wandering due to timing noise measured in young pulsars to date.
The spacing in $f$ is also held fixed for all pulsars at $\eta_f = 4 \times 10^{-10}\,\mathrm{Hz}$. 
This choice represents a trade-off between sensitivity and computational cost. 
When performing follow-up analysis of a glitch candidate, the range and spacing in $f$ are sometimes modified to encompass the pre- and post-glitch frequencies.
In the case of a large glitch, this means increasing the upper boundary of the $f$ region to a value on the order of $10^{-5}\,\mathrm{Hz}$. 
In this case the value of $\eta_f$ must also be increased to keep the total number of hidden states in the DOI small enough that the computation remains tractable.

The domain of interest and grid spacing in $\dot{f}$ vary between pulsars when searching for new glitches and setting upper limits.
Given a measured secular spindown $\dot{f}_\text{LS}$, we take the $\dot{f}$ region to be $\abs{\dot{f}} \leq \min(10^{-14}\,\mathrm{Hz}\,\mathrm{s}^{-1}, -0.1\dot{f}_\text{LS})$.
The region is empirically determined, guided by validation experiments with synthetic data, which show that timing-noise-driven excursions in $\dot{f}$ are small compared to $\dot{f}_\text{LS}$ \citep{melatosPulsarGlitchDetection2020}.
The $\dot{f}$ spacing $\eta_{\dot{f}}$ is chosen so that there are always 11 points in the domain.
This choice is motivated principally by a desire to keep computational cost under control.
As a side effect, pulsars with larger values of $\dot{f}_\text{LS}$ have larger values of $\sigma$ in the HMM's timing noise model: a larger $\dot{f}_\text{LS}$ gives a larger $\eta_{\dot{f}}$, and by equation (\ref{eqn:sigma_rot}) this in turns gives a larger $\sigma$.
While this relation between $\dot{f}_\text{LS}$ and $\sigma$ comports with the astrophysical fact that timing activity is correlated with $\dot{f}_\text{LS}$ \citep{arzoumanianTimingBehavior961994, hobbsAnalysisTimingIrregularities2010, lowerUTMOSTPulsarTiming2020}, it does not do so by design.
It is a consequence of pragmatic choices which aim to keep false alarms rare [in the case of equation (\ref{eqn:sigma_rot})] and computational cost low (in the case of the choice of $\eta_{\dot{f}}$).

\section{UTMOST glitches}
\label{sec:results}
We search for glitches in the UTMOST timing data in three stages.
In the initial stage, every pulsar is analysed using the HMM parameters set out in Table \ref{tab:hmm_params_initial}.
The results of the initial search are presented in Table \ref{tab:main_results}.
We report every glitch candidate with a Bayes factor greater than $K_\text{th} = 10^{1/2}$.
In the second stage, candidates are followed up with a simple veto procedure, and an estimates of the glitch parameters are calculated for those candidates which survive the veto.
%Each of these glitch candidates are followed up with a veto procedure, and the parameters of the candidates which survive the veto procedure are 

\begin{table*}
    \centering
    \caption{Properties of the initial glitch candidates detected in the first UTMOST data release.}
    \begin{tabular}{lccc}
        \hline
        Object & Epoch & $\ln K_1(k)$ & Vetoed?\\
        & MJD && Y/N \\\hline
        J0742$-$2822 & $57527 \pm 2$ & $4.05$ & Y \\
        J0835$-$4510 & $58218 \pm 4$ & $4.7 \times 10^4$ & N \\
        J1105$-$6107 & $57417 \pm 4$ & $1.27$ & Y \\
        J1257$-$1027 & $58651 \pm 10$ & $5.4$ & N \\
        J1359$-$6038 & $58189 \pm 2$ & $11.7$ & Y \\
        J1452$-$6036 & $58638 \pm 1$ & $126$ & N \\
        J1622$-$4950 & $58076 \pm 8$ & $8.46$ & N \\
        J1703$-$4851 & $58543 \pm 21$ & $47.5$ & N \\
        J1709$-$4429 & $58222 \pm 6$ & $1.17 \times 10^3$ & N \\
        J1731$-$4744 & $58007 \pm 2$ & $1.20 \times 10^5$ & N \\
        J1740$-$3015 & $58393 \pm 5$ & $4.47 \times 10^4$ & N \\
        \hline
    \end{tabular}
    \label{tab:main_results}
\end{table*}

Before estimating the parameters of each candidate, we check that the candidate is not due to a transient disturbance.
Such a disturbance may be caused by the conditions at the observatory, e.g. a clock error \citep{verbiestMeasurementUncertaintyPulsar2018}.
It may also have astrophysical origins.
Some pulsars exhibit ``mode-changing'', switching between a small number of distinct pulse profiles on timescales of minutes, which can lead to apparent jumps in the pulse phase \citep{backerPeculiarPulseBurst1970,helfandObservationsPulsarRadio1975a,wangPulsarNullingMode2007}.
Changes in propagation through the interstellar medium can also lead to similar apparent phase jumps \citep{goncharovIdentifyingMitigatingNoise2021}.
To exclude events of this kind, we re-run the HMM for each candidate using identical parameter choices, but with the ToAs immediately bracketing the candidate removed.
If one then obtains $K_1(k) < K_\text{th}$, the candidate is vetoed and no further analysis is performed.
The results of this veto procedure are noted in the right-most column of Table \ref{tab:main_results}.
Three candidates are vetoed in this way, leaving eight to be followed up in the second stage of the search.
This veto procedure does risk discarding candidates which correspond to true glitches, if the cadence around the candidate is low and the glitch is small.
Appendix \ref{apdx:vetoes} describes further investigation of each of the three vetoed candidates, in an effort to determine whether they are transient disturbances caused by one of the factors above.

The second stage of the search entails estimating the parameters of each candidate based on the sequence of most likely hidden states \citep{melatosPulsarGlitchDetection2020}, which gives the evolution of $f$, from which the approximate epoch and glitch size can be read off.
The follow-up analyses are performed first within an $f$ range which is wider than the range used in the initial analysis.
For all seven veto survivors which are ultimately identified as genuine glitch events (i.e. all but the candidate in J1622$-$4950, see Section \ref{subsec:results_J1622}), the maximum allowed frequency deviation is $2.5 \times 10^{-5}\,\mathrm{Hz}$ rather than $3 \times 10^{-7}\,\mathrm{Hz}$, so that large glitches are characterised more accurately.
The extended $f$ range degrades the sensitivity of the HMM to small glitches, as $\eta_f$ must increase to keep the computation tractable, which is why a smaller $f$ range is tested in the first stage.
After the analysis with an extended $f$ range is complete, each dataset is divided into pre- and post-glitch sections, and these sections are searched again with the smaller $f$ range and $\eta_f$ used in the initial search.
No additional glitch candidates are detected in this way.

The parameter estimation results are summarised in Table \ref{tab:full_analysis_results}.
\begin{table*}
    \centering
    \caption{Properties of the detected glitches confirmed by follow-up analyses. The fractional glitch sizes recovered in the HMM analysis are denoted by $\Delta f/f$, while the values reported by \citet{lowerUTMOSTPulsarTiming2020} are denoted by $(\Delta f/f)_\text{lit.}$. Phase ambiguity due to periodic observational scheduling prevents inferring $\Delta f/f$ for PSR J1452$-$6036 (see Section \ref{subsec:results_j1452}).}
    \begin{tabular}{ccccc}
        \hline
        Object & Epoch & $\Delta f/f$ & log Bayes factor & $(\Delta f/f)_\text{lit.}$\\
        & MJD & $\times 10^{-9}$ &  & $\times 10^{-9}$ \\\hline
        J0835-4510 & $57732 \pm 4$ & $1436 \pm 3$ & $7.43 \times 10^4$ & $1448^{+0.9}_{-0.8}$\\
        & $58521 \pm 7$ & $2467 \pm 13$ & $3.36 \times 10^3$ & $2501.2^{+2.6}_{-3.2}$ \\
        J1257$-$1027 & $58650\pm 16$ & $2.2 \pm 0.4$ & $16.4$ & $3.20^{+0.16}_{-0.57}$ \\
        J1452$-$6036 & $58606 \pm 3$ & -- & $1.40 \times 10^3$ & $270.7^{+0.3}_{-0.4}$ \\
        J1703$-$4851 & $58543 \pm 21$ & $10 \pm 2$ & $47.5$ & $19.0^{+1.0}_{-0.7}$ \\ 
        J1709$-$4429 & $58200 \pm 27$ & $2405 \pm 3$ & $220$ & $54.6 \pm 1.0$\\
        J1731$-$4744 & $58007 \pm 2$ & $3150 \pm 14$ & $1.40 \times 10^5$ & $3149^{+0.5}_{-0.4}$ \\
        J1740$-$3015 & $57476 \pm 17$ & $225 \pm 14$ & $1.43 \times 10^3$ & $237.7^{+13.2}_{-9.3}$ \\
        & $58240 \pm 11$ & $829 \pm 14$ & $9.37 \times 10^3$ & $842.3^{+7.1}_{-5.6}$\\
        \hline

    \end{tabular}
    \label{tab:full_analysis_results}
\end{table*}
Two outputs are of particular interest: the sequences of most likely frequency states\footnote{The point-wise estimate $\hat{f}(t_n)$ is the most likely value of $f$ at $t_n$ given the data $O_{1:N_T}$. This is subtly different from the $f$ component of the $n{\text{th}}$ element of the most likely sequence of hidden states.} $\hat{f}(t_n)$, and the posterior distribution $\gamma_f(t_n)$ of $f(t_n)$.
The most likely frequency states are obtained as the modes of the posterior distribution of states $q_i$ at each timestep $t_n$, where the posterior is denoted $\gamma_{q_i}(t_n)$ [defined in equation A13 of \citet{melatosPulsarGlitchDetection2020}].
The posterior frequency distribution $\gamma_f(t_n)$ is obtained by marginalising $\gamma_{q_i}(t_n)$ over $\dot{f}$.
Plots of $\hat{f}(t_n)$ and $\ln\gamma_f(t_n)$ for the follow-up analyses are shown in Figs. \ref{fig:J0835-4510_results}--\ref{fig:J1740-3015_results}.

We can obtain the posterior frequency derivative distribution in much the same way.
We do not present these distributions here, as coarse discretisation of $\dot{f}$ in the DOI often leads to unconstraining $\dot{f}$ posteriors.
An example is shown in Fig. \ref{fig:J1731-4744_fdot_posterior}, showing $\gamma_{\dot{f}}(t_n)$ from the follow-up analysis of J1731$-$4744 described in Section \ref{sec:results}.
The posterior has support over a significant fraction of the DOI, particularly after the glitch at the 103rd ToA gap, which makes it difficult to draw meaningful conclusions about the evolution of $\dot{f}$ across the glitch.

%We remind te reader that the current implementation of the HMM does not include quasi-exponential post-glitch recoveries in its phase model \citep{melatosPulsarGlitchDetection2020,dunnEffectsPeriodicityObservation2021}.
%Therefore one expects modest discrepancies between the glitch parameters (e.g. $\Delta f/f$) estimated by the HMM nd by previous authors, who do fit for post-glitch recoveries using traditional methods.
We remind the reader that the phase model of the HMM is not the same as the phase model of \textsc{tempo2}.
The HMM allows inter-glitch wandering of $f$ and $\dot{f}$, and transitions between the hidden $f$-$\dot{f}$ states occur only at the start of ToA gaps.
In addition, the HMM includes no explicit modelling of quasi-exponential post-glitch recovery processes.
For all of these reasons, we expect modest discrepancies between the glitch parameters estimated using the HMM and those reported by previous authors, who use \textsc{tempo2} and \textsc{temponest} to measure the glitch parameters.
Those glitches which merit additional discussion are covered in the remainder of this section.

\subsection{PSR J0835-4510}
\label{subsec:results_J0835}
PSR J0835-4510 (Vela) exhibits frequent large glitches, at a rate of roughly one every three years with $\Delta f/f \sim 10^{-6}$ typically \citep{howittNonparametricEstimationSize2018}.
The UTMOST data available for this pulsar consist of 1420 ToAs recorded between January 9 2014 and September 17 2018.
To keep the analysis computationally tractable, we divide the dataset into three sections of approximately 500 ToAs each, with the sections overlapping by 50 ToAs to ensure that any glitches that occur during the gaps between sections were not missed.
Details of the section boundaries can be found in Table \ref{tab:vela_sections}.
\begin{table}
    \centering
    \caption{Data segmentation in the PSR J0835$-$4510 analysis.}
    \begin{tabular}{ccc}
        \hline
        Section & Start MJD & End MJD \\\hline
        1 & 56666 & 56898 \\
        2 & 56868 & 57606 \\
        3 & 57552 & 58692\\\hline
    \end{tabular}
    \label{tab:vela_sections}
\end{table}
Plots of $\hat{f}(t_n)$ and $\ln\gamma_f(t_n)$ for the three sections are shown in Fig. \ref{fig:J0835-4510_results}.

We detected two large glitches in the data, both of which have been reported previously \citep{palfreymanGlitchObservedVela2016,sarkissianGlitchDetectedVela2019}.
The first, detected between MJD $57728$ and MJD $57734$, has $\Delta f/f = (1436 \pm 3)\times 10^{-9}$. 
The second, detected between MJD $58514$ and MJD $58529$, has $\Delta f/f = (2467 \pm 13)\times 10^{-9}$.
We note a feature which recurs several times throughout these analyses: $\gamma_f(t_n)$ shows multiple peaks for the timesteps following the second glitch, indicating the existence of multiple glitch models which describe the data well, despite being widely separated in frequency.
One can see this clearly in the bottom right panel for Fig. \ref{fig:J0835-4510_results}, where the yellow contour splits into three branches for ToA index $\geq 533$.
\citet{dunnEffectsPeriodicityObservation2021} showed that this effect is due to periodicity in the observation schedule.
If the separations between consecutive ToAs are nearly integer multiples of a common period $T$, there can be a degeneracy between glitch models with $\Delta f$ differing by $1/T$.
Since mid-2017 the Molonglo Observatory Synthesis Telescope  has operated as a transit instrument \citep{venkatramankrishnanUTMOSTSurveyMagnetars2020}, so each pulsar is observed at roughly the same local sidereal time for every observation.
Hence, to a good approximation, ToAs recorded by UTMOST after mid-2017 per pulsar are separated by integer numbers of sidereal days.
Indeed, the spacing between the peaks in $\gamma_f(t_n)$ in Fig. \ref{fig:J0835-4510_results} is close to $1/(1\,\text{sidereal day}) = 1.1606 \times 10^{-5}\,\mathrm{Hz}$.

The degeneracy between glitch models may be alleviated by additional observations which disrupt the periodic scheduling.
Fortunately, PSR J0835$-$4510 is an extremely well-studied pulsar, and the second glitch in the UTMOST dataset has been independently reported by several other facilities \citep{sarkissianGlitchDetectedVela2019,kerrFermiLATDetection2019, gancioUpgradedAntennasPulsar2020}.
\citet{kerrFermiLATDetection2019} estimated the size of the glitch to be $\Delta f/f = (2491.1 \pm 0.5) \times 10^{-9}$ based on data from the \emph{Fermi} Large Area Telescope \citep{atwoodLargeAreaTelescope2009}, and \citet{gancioUpgradedAntennasPulsar2020} estimated the size of the glitch to be $\Delta f/f = 2682 \times 10^{-9}$ based on observations taken at the Argentine Institute of Radio Astronomy.
These measurements are consistent with the HMM estimate, and with the estimate given by \citet{lowerUTMOSTPulsarTiming2020}.

\subsection{PSR J1452$-$6036}
\label{subsec:results_j1452}
We detected one glitch in this pulsar, occuring between MJD $58603.6$ and MJD $58604.6$.
The log Bayes factor over the no-glitch model is $1.4 \times 10^3$.

Plots of $\hat{f}(t_n)$ and $\ln\gamma_f(t_n)$ are shown in Fig. \ref{fig:J1452-6036_results}.
Inspection of $\gamma_f(t_n)$ indicates that caution is warranted when determining $\Delta f$ for this glitch.
The inferred $\hat{f}(t_n)$ shown in the left panel of Fig. \ref{fig:J1452-6036_results} suggests a glitch size of $\Delta f/f = 3869 \times 10^{-9}$.
However, as with PSR J0835$-$4510, the three peaks in $\gamma_f(t_n)$ indicate the existence of multiple glitch models which are widely separated in $\Delta f$ (the separation between peaks is approximately $1.1605 \times 10^{-5}\,\mathrm{Hz}$) but nevertheless describe the available data well.
This glitch was previously reported by \citet{lowerUTMOSTPulsarTiming2020} as occurring at MJD $58600.29(5)$ with $\Delta f/f = 270.7^{+0.3}_{-0.4}$.
Using the data in the UTMOST public data release, \citet{dunnEffectsPeriodicityObservation2021} demonstrated that the available data are consistent with $\Delta f/f = 270 \times 10^{-9} + N/(fT)$ with $N = 0,1,2$ and $T \approx 1\,\text{sidereal day}$, thereby including the \citet{lowerUTMOSTPulsarTiming2020} value as one possible option (with $N = 0$).
This result is in good agreement with the HMM analysis: the three peaks in the post-glitch frequency posterior generated by the HMM lie at $\Delta f/f = 269 \times 10^{-9}$, $2070\times 10^{-9}$, and $3869 \times 10^{-9}$.
Fortunately, independent observations at the Parkes radio telescope constrain the size of the glitch well, with \citet{JankowskiKeane2021} measuring $\Delta f/f = 270.52(3) \times 10^{-9}$.
This value is consistent with the \citet{lowerUTMOSTPulsarTiming2020} estimate and the smallest peak in $\gamma_f(t_n)$.
Note that we do not expect the tallest peak in the post-glitch frequency posterior to always correspond to the true glitch size when confounded by periodic scheduling \citep{dunnEffectsPeriodicityObservation2021}.

\subsection{PSR J1622$-$4950}
\label{subsec:results_J1622}
PSR J1622$-$4950 is a special object: it is a magnetar which shows large torque variations, with much larger variations in $\dot{f}$ than the DOI $\dot{f}_\text{LS} \pm 1 \times 10^{-14}\,\mathrm{Hz}\,\mathrm{s}^{-1}$ allowed in the initial search for glitches \citep{camiloRevivalMagnetarPSR2018}.
While this glitch candidate is not vetoed by removing ToAs either side of the glitch, it is probably an artifact caused by the first-pass DOI being too restrictive.
We note that the sequences of most likely $f$ and $\dot{f}$ states using the initial search parameters run up against the edges of the DOI.
We re-analyse the dataset using a DOI which is somewhat expanded in both $f$ and $\dot{f}$, with boundaries in $f$ at $f_\text{LS} \pm 5\times 10^{-6}\,\mathrm{Hz}$ and boundaries in $\dot{f}$ at $\dot{f}_\text{LS} \pm 1 \times 10^{-12} \,\mathrm{Hz}\,\mathrm{s}^{-1}$, searching for any glitch candidates in exactly the same way as before.
No glitch candidate is detected in this reanalysis, so we do not consider this candidate further.
For completeness, Figure \ref{fig:J1622-4950_results} shows the sequence of most likely frequencies and the posterior frequency probability for the re-analysis with the extended DOI.
The posterior has a relatively complex structure, because both the timing noise included in the HMM and the errors on individual TOAs are large, giving the HMM significant freedom in finding viable sequences of hidden states.
The large torque variations in PSR J1622$-$4950 make it difficult (though not impossible) to obtain a phase-connected timing solution covering timespans longer than a few months \citep{levinRadioloudMagnetarXray2010}.
We note briefly that the HMM offers a straightforward method of obtaining the pulse numbering and hence a phase-connected solution: from the sequence of most likely frequencies $\hat{f}(t_n)$ and frequency derivatives $\hat{\dot{f}}(t_n)$ it is easy to calculate the number of pulses during each gap via equation (\ref{eqn:hmm_accum_phase}).
From this information the relative pulse numbering is easily derived, and a phase-connected solution obtained.

\subsection{PSR J1709$-$4429}
We measured a glitch in this pulsar during the ToA gap between MJD $58172.9$ and MJD $58227.7$, with size $\Delta f/f = (2405 \pm 3) \times 10^{-9}$.
The log Bayes factor over the no-glitch model is $220$.

Plots of $\hat{f}(t_n)$ and $\ln\gamma_f(t_n)$ are shown in Fig. \ref{fig:J1709-4429_results}.
This detection corresponds to a glitch which was previously reported as occuring at MJD $58178 \pm 6$ with a glitch size of $\Delta f/f = 54.6 \pm 1.0 \times 10^{-9}$ \citep{lowerDetectionGlitchPulsar2018,lowerUTMOSTPulsarTiming2020}.
The glitch reported previously is smaller than the one we recover in this analysis.
As with the glitches in PSR J0835$-$4510 and PSR J1452$-$6036, the post-glitch frequency posterior is multiply peaked, with peaks separated by $\sim 1/(1\,\text{sidereal day})$ due to periodic observation scheduling.
\citet{dunnEffectsPeriodicityObservation2021} found that a combined dataset incorporating both the UTMOST data release and observations taken at the Parkes Observatory is consistent with $\Delta f/f = (2432\pm 0.1) \times 10^{-9}$.
Hence we expect that the glitch size recovered by the HMM reflects the true glitch properties.

\subsection{PSR J1740$-$3015}
\label{subsec:results_J1740}
In the follow-up analysis of PSR J1740$-$3015 we detected two glitches.
One glitch occurs in the ToA gap between MJD $57459$ and MJD $57486$, with size $\Delta f/f = (225 \pm 14)\times 10^{-9}$, and is detected with a log Bayes factor of $1.43 \times 10^3$.
The second glitch occurs in the ToA gap between MJD $58229$ and MJD $58243$, with size $\Delta f/f = (829\pm 14) \times 10^{-9}$, and is detected with a log Bayes factor of $9.37 \times 10^3$.
Plots of $\hat{f}(t_n)$ and $\ln\gamma_f(t_n)$ are shown in Fig. \ref{fig:J1740-3015_results}.

The first glitch was initially reported by \citet{jankowskiGlitchEventObserved2016} based on UTMOST data.
They reported $\Delta f/f = (227.29 \pm 0.03) \times 10^{-9}$, consistent with the HMM estimate.
Multiple peaks are visible in $\gamma_f(t_n)$ after the second glitch.
However, this second glitch was also reported by \citet{basuObservedGlitchesEight2020} based on data taken at the upgraded Giant Metrewave Telescope; they reported $\Delta f/f = 837.4(2) \times 10^{-9}$, consistent with the HMM analysis.
No post-glitch recovery term was included in their fit.

Two additional glitches have been reported in this pulsar during the timespan covered by the UTMOST first public release, occurring at MJD $57296.5 \pm 0.9$ and MJD $57346.0 \pm 0.6$ with sizes $\Delta f/f = 1.30 \pm 0.04 \times 10^{-9}$ and $\Delta f/f = 1.94 \pm 0.02 \times 10^{-9}$ respectively \citep{jankowskiGlitchEventObserved2015a,jankowskiGlitchEventObserved2016}.
We do not detect these glitches in our analysis.
This is not surprising, as the 90\% upper limit listed in Table \ref{tab:uls}, namely $\Delta f^{90\%}/f = 41 \times 10^{-9}$ (see Section \ref{sec:ul}), is an order of magnitude larger than the reported sizes of the undetected glitches.

\section{Size upper limits}
\label{sec:ul}
Having detected nine glitches among seven pulsars out of the 283 pulsars searched, we now turn to the question of completeness of this glitch sample.
In Section \ref{subsec:freq_ul} we discuss the 90\% frequentist upper limits set on 282 UTMOST pulsars, which are the main result of this section\footnote{We are unable to set a 90\% upper limit for the magnetar PSR J1622$-$4950, as discussed in Section \ref{subsec:freq_ul}.}.
In Section \ref{subsec:ul_pop_comp} we compare these upper limits to the observed population of glitches, and discuss the completeness of the sample of glitches reported in this work.
Finally in Section \ref{subsec:recovery_ul} we investigate how much quasi-exponential glitch recovery affects the upper limits of Section \ref{subsec:freq_ul}.

\subsection{Frequentist limits}
\label{subsec:freq_ul}
In order to assess completeness, we set 90\% frequentist upper limits $\Delta f^{90\%}$ on the sizes of undetected glitches for each pulsar in the UTMOST data release.
%The upper limits depend both on the intrinsic properties of the pulsar, and on extrinsic factors such as observing cadence and search parameters.
The upper limit is defined, such that there is a 90\% probability of detecting a glitch of size $\Delta f^{90\%}$ in the correct ToA gap, if the glitch occurs at a random epoch distributed uniformly over the entire dataset (excluding the first two and last two ToAs).
In Sections \ref{subsec:freq_ul} and \ref{subsec:ul_pop_comp}, for the sake of simplicity, we do not include a jump in $\dot{f}$ at the glitch epoch or an exponential post-glitch recovery; the latter effect is considered in Section \ref{subsec:recovery_ul}.
The probability of detection for a given pulsar and a given $\Delta f$ is estimated with 100 synthetic datasets generated using \textsc{libstempo} with a randomly chosen glitch epoch and noise injected at the level reported in the UTMOST data release.
The injected noise includes both Gaussian ToA measurement error and timing noise, e.g. spin wandering intrinsic to the pulsar \citep{goncharovIdentifyingMitigatingNoise2021}.
The procedures for generating the synthetic datasets and estimating $\Delta f^{90\%}$ are described in Appendix \ref{apdx:fake_data}.
The prescription for choosing HMM parameters is identical to the prescription for real data, as laid out in Section \ref{sec:method}.
A systematic upper limit injection study of this kind is practical only because the HMM runs fast and automatically without human intervention \citep{melatosPulsarGlitchDetection2020}.

\begin{figure}
    \centering
    \includegraphics[width=\columnwidth]{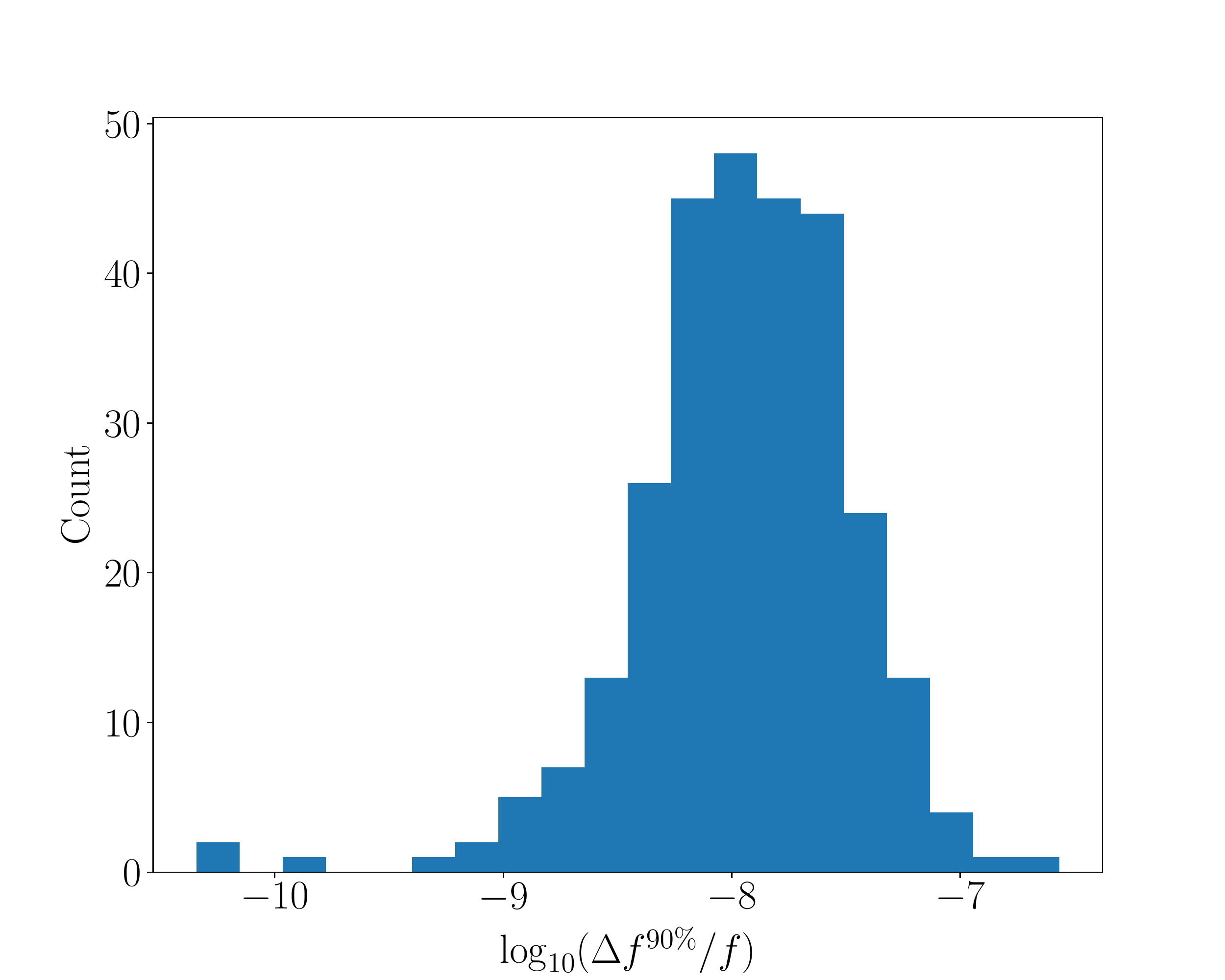}
    \caption{Histogram of 90\% frequentist upper limits on fractional glitch size for the 282 UTMOST pulsars analysed in this paper.}
    \label{fig:uls}
\end{figure}

Figure \ref{fig:uls} shows a histogram of $\Delta f^{90\%}/f$ for the 282 pulsars analysed here.
The majority (96\%) of the $\Delta f^{90\%}/f$ values lie between $10^{-9}$ and $10^{-7}$.
The mean fractional upper limit for our sample is $\langle \Delta f^{90\%}/f \rangle = 1.9 \times 10^{-8}$.
The minimum $\Delta f^{90\%}/f$ value is $4.5 \times 10^{-11}$ for the millisecond pulsar PSR J1730$-$2304, while the maximum is $2.7 \times 10^{-7}$ in the young pulsar PSR J1123$-$6259.
We do not list every value of $\Delta f^{90\%}/f$ here for readability, but Table \ref{tab:uls} lists the values for the seven pulsars in which we report at least one glitch in Section \ref{sec:results}.
A complete list of $\Delta f^{90\%}$ values can be found in the Supplementary Materials.
\begin{table}
    \centering
    \begin{tabular}{cc}
        \hline Object & $\Delta f^{90\%}/f$ \\
        & $\times 10^{-9}$ \\\hline
        J0835$-$4510 & $6.7$ \\
        J1257$-$1027 & $1.2$ \\
        J1452$-$6036 & $14$ \\
        J1703$-$4851 & $12$ \\
        J1709$-$4429 & $12$ \\
        J1731$-$4744 & $62$ \\
        J1740$-$3015 & $41$ \\
        \hline
    \end{tabular}
    \caption{90\% frequentist upper limits on the undetected glitch size in the seven UTMOST pulsars for which at least one glitch is detected by the HMM.}
    \label{tab:uls}
\end{table}

No upper limit is obtained for PSR J1622$-$4950, where strong timing noise in the HMM (see Section \ref{subsec:results_J1622}) and a periodic observation schedule mean that it is not possible to attain a detection probability of 90\% for any plausible glitch size.
Typically, even in the presence of significant timing noise one can increase $\Delta f$ to a point where timing noise can no longer account for the frequency jump.
However, if observations are periodic as they are here, with a period of $1\,\text{sidereal day}$, then glitches with sizes $\Delta f$ larger than $N/(1\,\text{sidereal day})$ (where $N$ is an integer) are recovered as glitches with size $\Delta f - N/(1\,\text{sidereal day})$.
Hence they may never be detected by the HMM if the timing noise is large enough to account for a change in frequency of $\Delta f - N/(1\,\text{sidereal day})$.

For the data analysed here, the sensitivity of the HMM is principally controlled by the largest observing gaps in the data.
Large gaps allow for significant deviations in the spin frequency to be absorbed into the timing noise model in the HMM, with the expected fractional upper limit proportional to $\dot{f}_+ \max_n x_n$.
When this effect dominates, a regular and frequent observing cadence is most useful in obtaining more stringent upper limits.
However, we caution that this is not true in all regimes.
When the observing cadence is short enough, or the allowed wandering due to timing noise is small enough, the sensitivity of the HMM is instead controlled by the phase uncertainty incorporated into the HMM via the $\kappa$ parameter (see Section \ref{subsec:trans_emis}).
In this case the expected fractional upper limit is roughly proportional to $2\pi\kappa^{-1/2}\langle x_n\rangle^{-1}$, and increasing the cadence (decreasing $\langle x_n \rangle$) further will give \emph{lower} sensitivity, if $\kappa$ is dominated by the contribution from $\sigma_{\text{ToA}}$, which is independent of $x_n$.
We refer the reader to section 6 and appendix G of \citet{melatosPulsarGlitchDetection2020} for further discussion on the sensitivity of the HMM glitch detector.

\subsection{Population-level comparison}
\label{subsec:ul_pop_comp}
We compare the upper limits obtained in Section \ref{subsec:freq_ul} to the observed size distribution aggregated across the entire pulsar population, as recorded in the Jodrell Bank Observatory (JBO) glitch catalogue\footnote{\href{http://www.jb.man.ac.uk/pulsar/glitches/gTable.html}{http://www.jb.man.ac.uk/pulsar/glitches/gTable.html}} \citep{espinozaStudy315Glitches2011}.
This is not exactly a like-for-like comparison: the catalogue of observed glitches combines a wide variety of datasets and analyses, with varying observation scheduling and glitch detection strategies.
Nevertheless it is instructive to ask what categories (if any) of glitches observed in other pulsars are not detectable by the HMM in the UTMOST data release.

\begin{figure}
    \centering
    \includegraphics[width=\columnwidth]{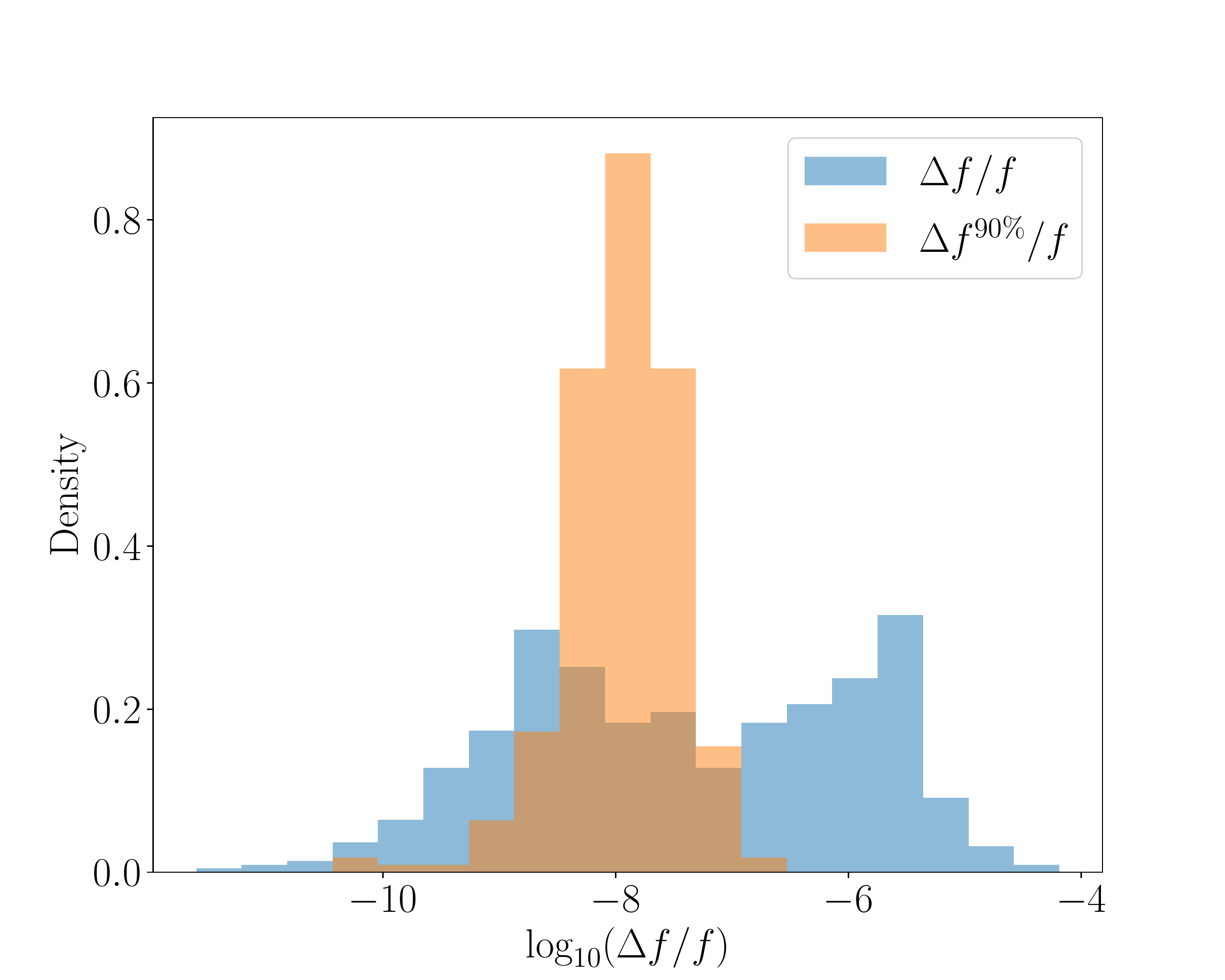}
    \caption{Histogram of size upper limits $\Delta f^{90\%}/f$ computed in Section \ref{subsec:freq_ul} (orange) compared with sizes $\Delta f/f$ detected in the entire pulsar population in the Jodrell Bank Observatory catalogue (blue).}
    \label{fig:ul_obs_comparison}
\end{figure}

Fig. \ref{fig:ul_obs_comparison} shows a histogram of all $\Delta f/f$ values listed in the JBO glitch catalogue overlaid with a histogram of the upper limits obtained in this analysis.
There is a population of detected glitches in the JBO catalogue with $10^{-12} \lesssim \Delta f/f \lesssim 10^{-9}$ which are smaller than the 90\% upper limits obtained for most of the pulsars in our sample.
Glitches in this size range are unlikely to be detected by the search in this paper.
Of course, this raises the interesting question of whether some glitches in the JBO catalogue with $\Delta f/f \lesssim 10^{-9}$ are false alarms.
This comes down to distinguishing timing noise from glitches through Bayesian model selection and calculating $\Delta f^{90\%}$ for the relevant observational studies in the literature, a task which is challenging without an unsupervised algorithm like the HMM \citep{janssen30GlitchesSlow2006,chukwudeObservationsMicroglitchesHartebeesthoek2010a, espinozaNeutronStarGlitches2014, yuDetectionProbabilityNeutron2017}.

\iffalse
The variation in upper limits between pulsars is linked closely to the strength of the timing noise included in the HMM, parametrised by $\sigma$.
The prescription for choosing $\sigma$ on a per-pulsar basis is set out in Section \ref{subsec:trans_emis}: it is a conservative rule of thumb.
In the absence of a complete means of both characterising the timing noise and incorporating an appropriate timing noise model into the HMM, the rule of thumb serves as a compromise which does not require by-hand adjustment of the HMM for every pulsar analysed.
However, this compromise leads to upper limits which for many pulsars do not approach the intrinsic upper limit of detectability (that is, the upper limit of detectability for a perfectly tuned HMM).
\fi

By way of comparison, we briefly highlight three other investigations of glitch detectability which are similar in spirit to the current work.
\citet{janssen30GlitchesSlow2006} performed Monte Carlo injections for a single pulsar, PSR J0358$+$5413, finding that glitches as small as $\Delta f/f = 10^{-11}$ can be detected by eye.
It is unclear, however, how confidently such glitches can be detected.
\citet{espinozaNeutronStarGlitches2014} employed an automated glitch detection algorithm to search for glitches in the Crab pulsar, and reported a minimum glitch size in that case, which is intrinsic to the pulsar and not an artifact of the detector performance.
This technique has also been applied to the Vela pulsar, and a lack of small glitches was reported in that case also \citep{EspinozaAntonopoulou2021}.
However, the detector returns a large number of (anti-)glitch candidates which must be classified after the fact as timing noise or glitches.
This obstructs the characterisation of the detector in a controlled environment via a suite of synthetic data tests, as well as the application of this technique to a large number of datasets.
\citet{yuDetectionProbabilityNeutron2017} assessed the completeness of the glitch catalogue reported in \citet{yuDetection107Glitches2013} through Monte Carlo simulations in which glitch detection was performed using \textsc{temponest}.
They concluded that the reported glitch catalogue contains all glitches detectable by manual inspection of timing residuals.
However the criterion for a positive detection relies on knowing the true glitch epoch; it cannot be extended to finding previously unknown glitches.
We emphasise that although some glitches may be missed, the upper limits calculated here are derived from simulated searches of every pulsar individually.
%Crucially, the search procedure used to set the upper limits is the same as the search procedure used when looking for unknown glitches.

\subsection{Post-glitch recovery}
\label{subsec:recovery_ul}
Many (but not all) glitches exhibit a degree of recovery over timescales of days to months, such that part (or all) of $\Delta f$ reverses, leaving a permanent frequency jump $\Delta f_\text{p}$ \citep{shemarObservationsPulsarGlitches1996}.
The recovery is typically modelled as one or more exponential terms in the post-glitch frequency evolution, viz. \begin{equation} f(t) = f(0) + \Delta f_\mathrm{p} + \sum_{i=1}^M \Delta f_i e^{-t/\tau_i} \end{equation} for a hypothetical glitch occurring at $t = 0$, where the $\Delta f_i$ are the sizes of the $M$ exponentially recovering components, with recovery timescales $\tau_i$.
In many events one has $M=1$, but where the pulsar is well-observed following the glitch, more exponential terms may be incorporated, e.g. $M \leq 4$ \citep{dodsonHighTimeResolution2002}.
The glitch population as a whole exhibits a wide variety of recovery behaviour.
Values of the healing parameter \begin{equation} Q = \frac{\sum_i \Delta f_i}{\Delta f_\mathrm{p} + \sum_i \Delta f_i} \end{equation} are typically between $0$ and $1$, with $Q \ll 1$ being more common for large glitches ($\Delta f_\mathrm{p}/f \gtrsim 10^{-6}$) \citep{yuDetection107Glitches2013}.

\begin{figure}
    \centering
    \includegraphics[width=\columnwidth]{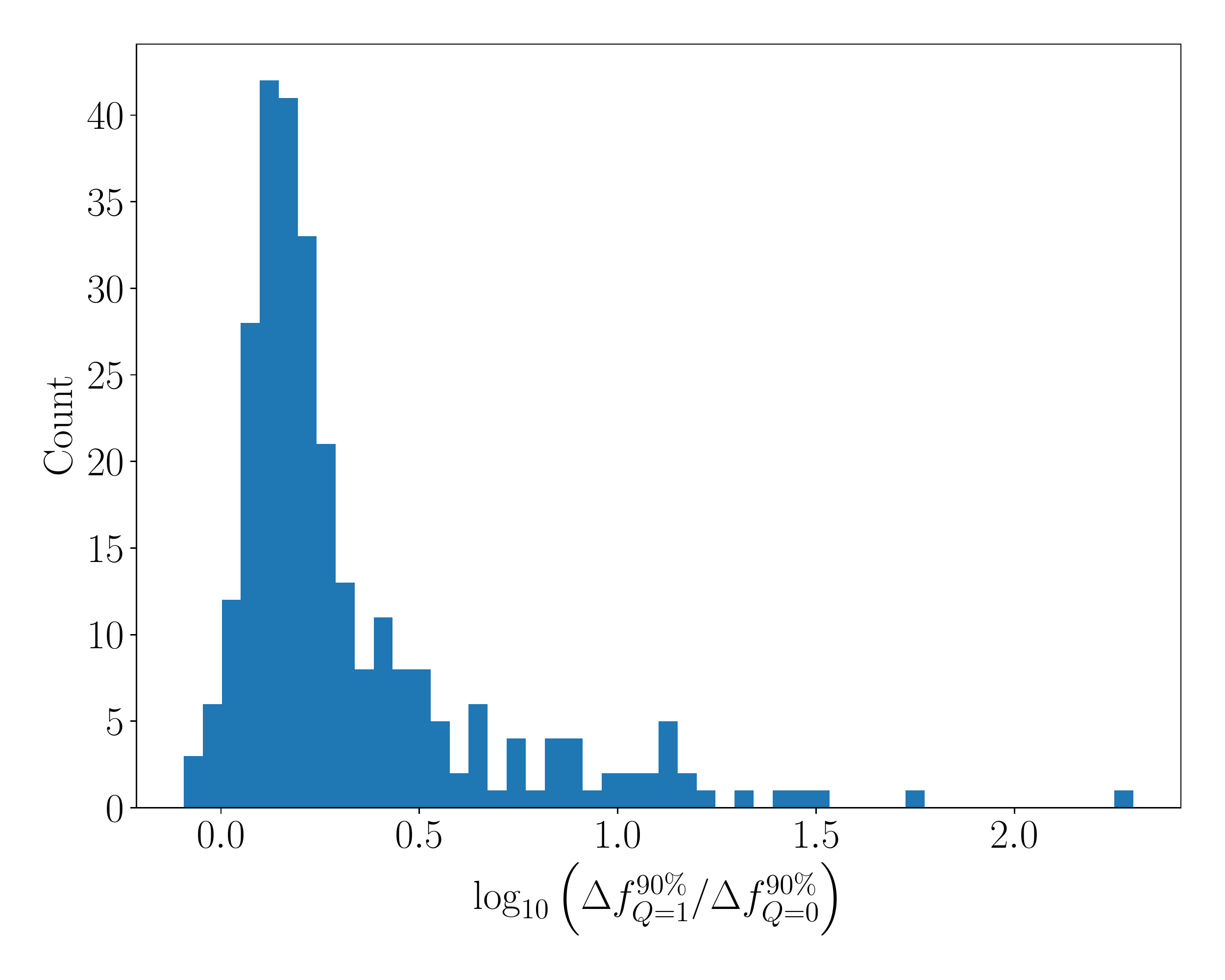}
    \caption{Histogram for 282 UTMOST pulsars of the ratio of 90\% glitch size upper limit computed with a completely recovering glitch ($\Delta f^{90\%}_{Q=1}$) to the 90\% upper limit computed with no recovery ($\Delta f^{90\%}_{Q=0}$).}
    \label{fig:ul_recovery_ratio}
\end{figure}

In Section \ref{subsec:freq_ul} we set 90\% frequentist upper limits on the sizes of undetected glitches assuming a glitch model (both in the simulated data and the HMM) with no recovery (i.e. $Q = 0$).
We now investigate whether including recovery in the simulated data significantly changes these upper limits.
To this end, we recompute 90\% frequentist upper limits as in Section \ref{subsec:freq_ul}, but now assume that the glitch recovers completely (i.e. $Q = 1$ in the simulated data) on a timescale of $\tau_1 = 100\,\mathrm{d}$, typical of many pulsars \citep{yuDetection107Glitches2013}.
Note that we do not modify the phase model of the HMM in any way --- no attempt is made to model the exponential recovery as part of the glitch detection step.
Fig. \ref{fig:ul_recovery_ratio} shows a histogram of the ratios between the $Q = 1$ and $Q = 0$ values of $\Delta f^{90\%}$ for each pulsar.
In the majority of cases the effect of recovery is not severe: for 87\% of the pulsars one obtains $\Delta f^{90\%}_{Q=1} < 5\Delta f^{90\%}_{Q=0}$.
Hence we do not expect complete glitch recovery on month-long timescales to affect significantly the results presented in Section \ref{subsec:freq_ul}.
The outliers with large ratios $\Delta f^{90\%}_{Q=1}/\Delta f^{90\%}_{Q=0}$ typically have small intrinsic $\dot{f}_\text{LS}$, with $\dot{f}_\text{LS} > -10^{-15}\,\mathrm{Hz}\,\mathrm{s}^{-1}$.
In this regime the extra $\dot{f}$ due to quasi-exponential recovery is much larger than the range of $\dot{f}$ in the DOI, which covers only $\pm 0.1\dot{f}_\text{LS}$ (see Section \ref{subsec:doi}).
Hence the HMM struggles to track the evolution of $f$ and $\dot{f}$, and is correspondingly less sensitive.
Conversely, for six pulsars we have $0.8 < \Delta f^{90\%}_{Q=1}/\Delta f^{90\%}_{Q=0} < 1$.
In all but one case these pulsars are monitored with relatively high cadence, and have sufficiently wide ranges in the $\dot{f}$ DOI to allow the tracking of the extra $\dot{f}$ from the exponential recovery\footnote{In the remaining case, which is PSR J1736$-$2457, we expect that simple statistical fluctuation is the cause: the observing cadence is relatively low, with several gaps of $10$--$60$ days present in the data. In this case, when relaxations are included and the effect of a glitch on $f$ and $\dot{f}$ can decay away within one or two post-glitch ToAs, the estimated value of $\Delta f^{90\%}$ may depend somewhat on the epochs of the injected glitches in the synthetic datasets.}
Hence we expect that the change in $\dot{f}$ from the recovery allows the HMM to detect glitches more readily in this regime.

The above analysis only considers one point in the space of possible $(Q, \tau_1)$ choices (though it is a fairly typical point).
A full exploration of the $Q$-$\tau_1$ plane for every pulsar in the UTMOST data release is beyond the scope of this paper.
We emphasise that the conclusions drawn about upper limits here and in Section \ref{subsec:freq_ul} are conditional not only on the analysis method but also on the assumed glitch model, a feature of any timing analysis.

\section{Conclusion}
In this paper we present a search for glitches using a HMM in 283 pulsar timing datasets released by the UTMOST pulsar timing programme, covering observations taken between October 2015 and August 2019.
We detect nine glitches among seven pulsars, all of which have been previously reported.
%The inferred properties of the detected glitches are consistent with previously reported measurements, although we note some complications arising from periodicity in the UTMOST observation schedule, which prevents an accurate recovery of one glitch, in PSR J1452$-$6036 based on UTMOST data alone.
The inferred $\Delta f$ is usually consistent with previous discoveries, except when there is ambiguity due to near-periodic scheduling \citep{dunnEffectsPeriodicityObservation2021}.
In this case the discrepancy in $\Delta f$ can be large, viz. an integer multiple of $T^{-1}$, where $T$ is the observation scheduling period.
For all the glitches detected in this work, complementary observations by other observatories allow the unambiguous determination of $\Delta f$.
In principle the $\Delta f$ inferred from an HMM analysis may be biased by quasi-exponential post-glitch recovery, which is not included in the HMM in its current implementation.
However, this effect is demonstrated to be small for the glitches measured in this work, typically $\lesssim 2\%$ for the $7$ objects studied here.
Incorporating post-glitch recoveries into the HMM (at the expense of introducing new parameters) is a priority for future work.

For each object, we perform injection studies to set frequentist upper limits on the size of undetected glitches.
The mean 90\% upper limit on the fractional size of undetected glitches is $\langle\Delta f^{90\%}/f\rangle = 1.9 \times 10^{-8}$.
The smallest value of $\Delta f^{90\%}/f$ is $4.1 \times 10^{-11}$, calculated for the millisecond pulsar PSR J1730-2304.
The largest value of $\Delta f^{90\%}/f$ is $2.7 \times 10^{-7}$, calculated for the young pulsar PSR J1123$-$6259.
Obtaining more stringent upper limits using the existing data would require a more complete characterisation of the timing noise and a comprehensive understanding of how different timing noise models ought to be included in the HMM, a challenge which is faced by other glitch detection schemes \citep{chukwudeObservationsMicroglitchesHartebeesthoek2010a, espinozaNeutronStarGlitches2014,SinghaBasu2021}.
Independent of these timing noise considerations, future observing campaigns can provide stricter upper limits with higher cadence and/or more sensitive observations.
We show that glitch recovery has a mild effect on the upper limits; the upper limits for 87\% of the objects increase by no more than a factor of $5$, if it is assumed that the undetected glitch recovers completely on a fiducial time-scale of $100\,\mathrm{d}$.

Understanding the completeness of glitch catalogues is essential to falsifying models of glitching behaviour. 
For instance, \citet{melatosSizewaitingtimeCorrelationsPulsar2018} predicted (under certain weak, astrophysics-independent assumptions) that pulsars with large values of $-\dot{f}\langle\Delta t\rangle$ (where $\langle\Delta t\rangle$ is the mean waiting time between glitches) should show significant correlations between the size of a glitch and the forward waiting time to the next glitch.
Similarly, pulsars with small values of $-\dot{f}\langle\Delta t\rangle$ are predicted to be the most likely to exhibit correlations between glitch size and backward waiting time, although the latter correlations are predicted to be weaker.
As a foretaste of what is possible, Table \ref{tbl:fdot_deltat} shows the five highest and five lowest values of $-\dot{f}\langle\Delta t\rangle$ amongst the pulsars which have been observed in the first UTMOST data release, as well as the Pearson corrrelation coefficients between $\Delta f/f$ and the forward and backward waiting times.
While most of these pulsars do not yet have enough glitches observed for any statistically significant conclusions to be drawn, there is tentative evidence for strong forward correlations in the pulsars with the largest values of $-\dot{f}\langle\Delta t\rangle$.
Continued high-cadence monitoring of these pulsars and a good understanding of the likelihood that a glitch of a given size might not have been detected are essential to falsifying the proposed relations.
\begin{table}
    \centering
    \caption{Top and bottom five $-\dot{f}\langle\Delta t\rangle$ values among pulsars observed by UTMOST. The Pearson correlation coefficients between $\Delta f/f$ and the forward ($r_+$) and backward ($r_-$) waiting times are also listed.\label{tbl:fdot_deltat}. The objects in the top (bottom) half of the table are more likely to exhibit cross-correlations between glitch size and forward (backward) waiting time.}
    \begin{tabular}{lcccr}\hline
        Object & $N_\text{g}$ & $-\dot{f}\langle\Delta t\rangle$ (Hz) & $r_+$ & $r_-$ \\\hline
        J1709$-$4429 & 5 & $1.8 \times 10^{-3}$ & $0.66$ & $-0.47$\\
        J1803$-$2137 & 6 & $1.4 \times 10^{-3}$ & $0.91$ & $-0.24$\\
        J0835$-$4510 & 20 & $1.2 \times 10^{-3}$ & $0.24$ & $0.55$\\
        J1048$-$5832 & 6 & $6.0 \times 10^{-4}$ & $0.58$ & $-0.48$\\
        J1105$-$6107 & 5 & $4.2 \times 10^{-4}$ & $0.87$ & $-0.37$\\\hline
        J1731$-$4744 & 5 & $4.4 \times 10^{-5}$ & $-0.71$ & $0.99$\\
        J1740$-$3015 & 36 & $3.5 \times 10^{-5}$ & $0.29$ & $-0.02$\\
        J1705$-$1906 & 4 & $1.2 \times 10^{-5}$ & $0.97$ & $-0.54$\\
        J1825$-$0935 & 7 & $1.1 \times 10^{-5}$ & $0.91$ & $-0.30$\\
        J1902$+$0615 & 6 & $2.0 \times 10^{-6}$ & $0.49$ & $-0.31$\\
        \hline
    \end{tabular}
\end{table}
Improved understanding of the completeness of glitch catalogues is also important to studies of the physical conditions involved.
Several authors have investigated the nature of the neutron superfluid in the inner crust by studying glitching behaviour (e.g. \citealp{anderssonPulsarGlitchesCrust2012,hoPinningSuperfluidMeasuring2015,montoliRoleMassEquation2020,montoliStatisticalEstimatesPulsar2021}).
The quantity of interest is frequently the cumulative fractional change in spin frequency due to glitches, $\mathcal{A} \propto \sum_i \Delta f_i/f$. 
An understanding of the completeness of the glitch sample is essential to understanding the uncertainty on $\mathcal{A}$, and by extension understanding the implications of measured values of $\mathcal{A}$ on the underlying physics. 

\section*{Acknowledgements}
The authors are most grateful to the UTMOST team for their hard work in collecting the data which forms the first UTMOST open data release.
The Molonglo Observatory is owned and operated by the University of Sydney with support from the School of Physics and the University.
Parts of this research are supported by the Australian Research Council (ARC) Centre of Excellence for Gravitational Wave Discovery (OzGrav) (project number CE170100004) and ARC Discovery Project DP170103625.
L. Dunn is supported by an Australian Government Research Training Program Scholarship and by the Rowden White Scholarship.
This work was performed on the OzSTAR national facility at Swinburne University of Technology. The OzSTAR program receives funding in part from the Astronomy National Collaborative Research Infrastructure Strategy (NCRIS) allocation provided by the Australian Government.

\section*{Data availability}
{The public UTMOST data underlying} this work are available at \url{https://github.com/Molonglo/TimingDataRelease1/}.
%\href{https://github.com/Molonglo/TimingDataRelease1/}{https://github.com/Molonglo/TimingDataRelease1/}.
The other data underlying this work will be shared on reasonable request to the corresponding author.

%%%%%%%%%%%%%%%%%%%%%%%%%%%%%%%%%%%%%%%%%%%%%%%%%%
\bibliography{main}{}
\bibliographystyle{mnras}
%%%%%%%%%%%%%%%%% APPENDICES %%%%%%%%%%%%%%%%%%%%%
\appendix

\section{Follow-up analysis of vetoed candidates}
\label{apdx:vetoes}
In Section \ref{sec:results} we describe the vetoing of three candidates.
Given that these vetos proceed by removing the ToAs bracketing each of the candidates, there is a chance that genuine glitch events may be discarded by this procedure, if the glitch is close to the limit of detectability.
In this appendix we investigate each of the vetoed candidates in more detail, with the aim of clarifying their origin.

\subsection{PSR J0742$-$2822}
The case of PSR J0742$-$2822 turns out to be straightforward.
The dispersion measure (DM) in the ephemeris file provided in the UTMOST data release was mistakenly quoted as $681\,\mathrm{pc}\,\mathrm{cm}^{-3}$, whereas the correct value is $74\,\mathrm{pc}\,\mathrm{cm}^{-3}$.
A large error in the DM can produce significant scatter in the timing residuals, as even small variations in the central frequency of each observation lead to large corrections to the ToAs.
After correcting the DM in the ephemeris, we find that the scatter in the residuals in the vicinity of the candidate is reduced by a factor of 20.
Finally, we re-run the HM analysis with the updated DM value and find no candidate.
Hence we reject the initial candidate as non-astrophysical.

\subsection{PSR J1105$-$6107}
Inspection of the timing residuals surrounding the candidate in PSR J1105$-$6107 reveals no obvious features, and inspection of the raw archives similarly reveals no disturbance in the vicinity of the candidate.
As a consistency check, we re-generate the ToAs for PSR J1105$-$6107 using \textsc{psrchive} \citep{HotanvanStraten2004} and re-run the HMM analysis.
No candidate is returned, and we thus reject the candidate as non-astrophysical.

We are unable to identify a clear reason for the discrepancy between the ToAs in the data release and the re-generated set.
As we are aware, both the archives containing the folded observations and the standard profile used to generate the times of arrival have not changed between the UTMOST data release and the re-generation described here.
However, records of the UTMOST data release preparation are not sufficiently detailed to allow us to check this, and it is possible that the profile used to generate the ToAs in the UTMOST data release was not optimal (e.g. not sufficiently smoothed).

\iffalse
\subsection{PSRs J0742$-$2822 and J1105$-$6107}
We group these candidates together as the underlying cause appears to be same in both cases.
Inspection of the timing residuals surrounding each candidate reveals no obvious features, and inspection of the raw archives similarly reveals no disturbances in the vicinity of the candidates.

As a consistency check, we re-generate the ToAs for the entire datasets for both pulsars using \textsc{psrchive} \citep{HotanvanStraten2004} and re-run the HMM analysis.
No candidates are returned in these re-analyses.
The reason for the difference 
We note that the residuals from these re-generated ToAs exhibit significantly lower scatter, by approximately a factor of 20 in the case of PSR J0742$-$2822, and a factor of 2 in the case of PSR J1105$-$6107.
We are unable to identify a clear reason for the excess scatter in the ToAs released in the public data release -- the data and standard profile used to re-generate the ToAs are identical to those used to generate the ToAs in the public data release.
Despite being unable to identify a clear root cause in these cases, given that both candidates disappear when the newly generated ToAs are analysed instead we are confident that these candidates do not correspond to true glitch events.
\fi

\subsection{PSR J1359$-$6038}
Inspection of the timing residuals surrounding the candidate in PSR J1359$-$6038 reveals that a single ToA at MJD $58190.7$ is displaced away from the rest of the surrounding ToAs by approximately $0.4\,\mathrm{ms}$.
This is significant compared to the uncertainty on this ToA of $60\,\mu\mathrm{s}$.
To check whether this displaced ToA is due to conditions at the observatory, we inspect the timing residuals of other pulsars that were observed no more than twelve hours before or after the ToA in question.
We identify multiple pulsars in which the observation nearest MJD $58190.7$ is displaced by approximately the same amount in the same direction, for example PSRs J1146$-$6030, J1600$-$3053\footnote{In the case of PSR J1600$-$3053 the displaced ToA was removed manually during the preparation of the public data release.}, and J1644$-$4559.
Thus we conclude that the candidate in PSR J1359$-$6038 is due to local conditions at the observatory and has no astrophysical origin.

\section{Synthetic dataset generation and upper limit estimation}
\label{apdx:fake_data}
In order to set frequentist upper limits we perform injection studies for each pulsar.
We first outline the procedure for generating a single synthetic dataset for one object, which is based on a given UTMOST dataset (i.e. with identical ephemeris, observing cadence, and ToA uncertainties), with a glitch of size $\Delta f$ injected.
\begin{enumerate}
    \item A glitch epoch $t_\text{g}$ is chosen at random, uniformly distributed between the second and second-last ToAs. Glitches which occur in either the first or last ToA gap are indistinguishable from a single outlier ToA due to some external factor, so we do not consider them when setting upper limits here.
    \item A new phase model is generated by \textsc{libstempo} which matches the UTMOST phase model, except that a glitch term $\Delta\phi_\text{g}(t) = \Theta(t-t_\text{g})\Delta f (t-t_\text{g})$ is added [where $\Theta(t)$ is the Heaviside step function].
    \item Using the ToAs of the original dataset as a starting point, a new set of ToAs is generated by shifting the original ToAs slightly so that they show zero residuals with respect to the new phase model.
    \item Noise is introduced into the new set of ToAs at the levels reported in the UTMOST data release. We use the \texttt{add\_efac}, \texttt{add\_equad}, and \texttt{add\_rednoise} functions in \textsc{libstempo}, using the EFAC, EQUAD, and red noise parameters reported for each pulsar in the UTMOST data release.
    \item The new phase model and new set of ToAs are saved as a synthetic dataset.
\end{enumerate}
This procedure ensures that the synthetic datasets closely match the true datasets in various important aspects, e.g. basic timing model parameters, observing cadence, and noise characteristics.

The procedure for estimating the probablity of detection for a glitch size $\Delta f$ [denoted $P_\text{d}(\Delta f)$] in a single pulsar is straightforward:
\begin{enumerate}
    \item Generate $100$ synthetic datasets with a glitch of size $\Delta f$ injected, according to the procedure in the paragraph above.
    \item Analyse each dataset with the HMM following the method outlined in Section \ref{sec:method}.
    \item For each dataset, determine the Bayes factor $K$ between the model $M_1(k_\text{inj})$ with a glitch included in the ToA gap indexed by $k_\text{inj}$ corresponding to the injected glitch epoch and the model $M_0$ with no glitch included.
    \item The proportion of synthetic datasets with $K > K_\text{th}$ gives an estimate of $P_\text{d}(\Delta f)$.
\end{enumerate}

Finally we give a simple prescription for estimating the value of $\Delta f^{90\%}$ from $P_\text{d}(\Delta f^{90\%}) = 0.9$, i.e. the $90\%$ upper limit on the size of undetected glitches in each pulsar.

\begin{enumerate}
    \item Choose the starting range of glitch sizes to be $[\Delta f_-, \Delta f_+] = [10^{-9}, 10^{-6}]\,\mathrm{Hz}$.
    \item Choose a glitch size $\Delta f$ by bisecting the range logarithmically, i.e., \begin{equation} \log_{10}(\Delta f) = \log_{10}(\Delta f_-) + [\log_{10}(\Delta f_+) - \log_{10}(\Delta f_-)]/2 \label{eqn:ul_freq_choice}\end{equation} where all the frequencies are understood to be in units of Hz.
    \item Calculate $P_\text{d}(\Delta f)$ as outlined previously.
    \item If $\abs{P_\text{d}(\Delta f) -0.9} \leq 0.01$, terminate and take $\Delta f$ as the 90\% frequentist upper limit $\Delta f^{90\%}$.
    \item Otherwise, revise the glitch size range as follows:
    \begin{enumerate}
        \item If $P_\text{d}(\Delta f) > 0.9$, set $\Delta f_+ = \Delta f$.
        \item If $P_\text{d}(\Delta f) < 0.9$, set $\Delta f_- = \Delta f$.
    \end{enumerate}
    \item Return to step (ii).
\end{enumerate}
This is essentially a binary search over possible upper limits.

\section{Posterior distributions and frequency tracks for HMM analyses}
\label{apdx:glitch_plots}
\begin{figure*}
    \centering
    \includegraphics[width=\textwidth]{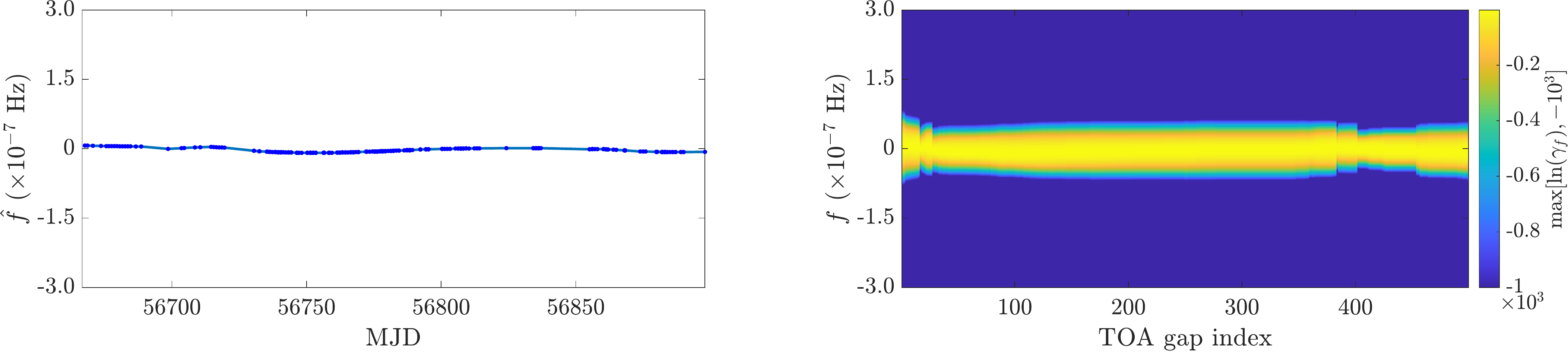}\\
    \includegraphics[width=\textwidth]{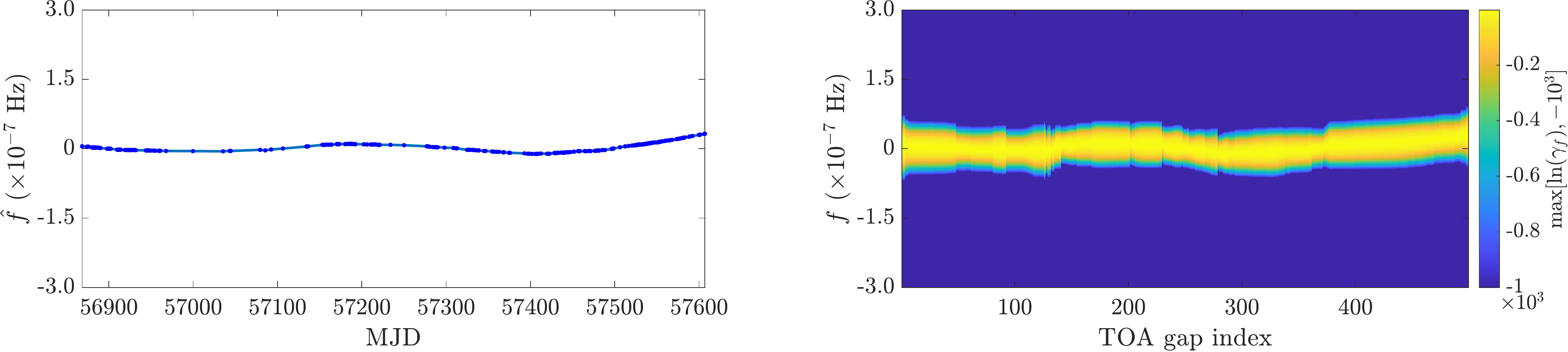}\\
    \includegraphics[width=\textwidth]{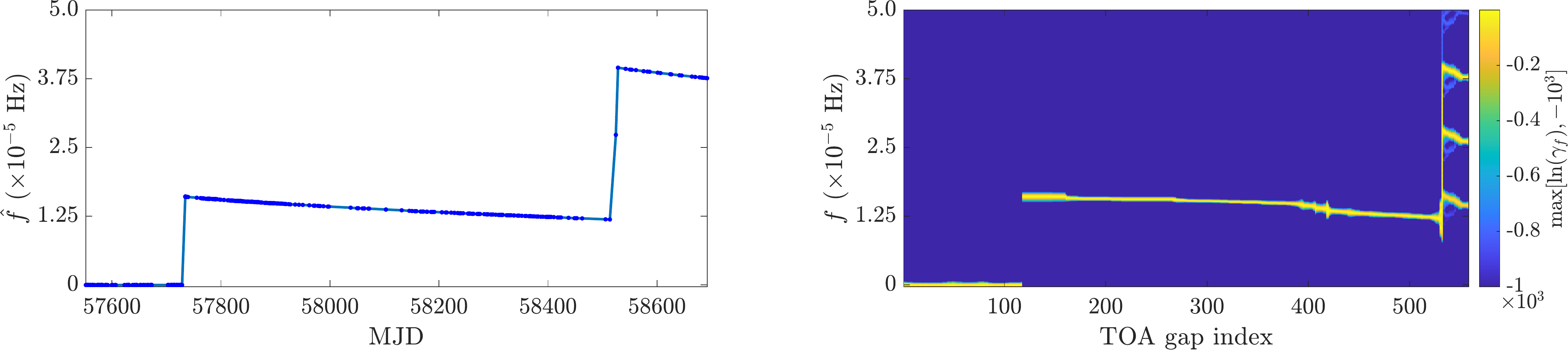}
    \caption{Sequence of most likely frequencies $\hat{f}(t_n)$ \emph{(left)} and heatmap of posterior frequency probability $\ln[\gamma_f(t_n)]$ \emph{(right)} for the HMM follow-up analysis of PSR J0835$-$4510.
    Frequency is on the vertical axis in all panels, and the range of the vertical axis is the full range of the DOI.
    Note that the horizontal axes for the two panels are not exactly the same: the left panels have MJD on the horizontal axis, while the right panels have ToA gap index on the horizontal axis instead, for ease of plotting. 
    The three rows correspond to sections 1, 2 and 3 from top to bottom as described in Table \ref{tab:vela_sections}.}
    \label{fig:J0835-4510_results}
\end{figure*}

\begin{figure*}
    \centering
    \includegraphics[width=\textwidth]{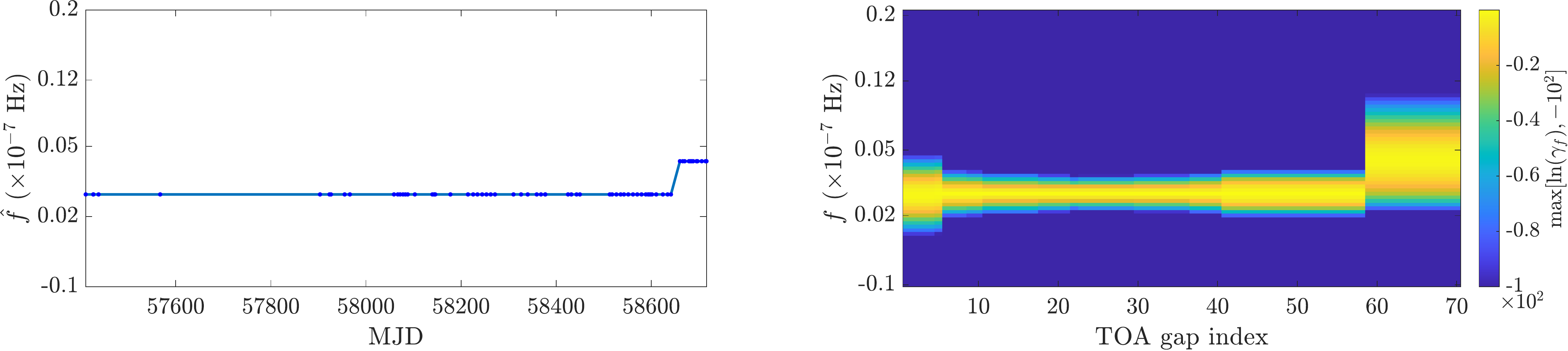}
    \caption{As in Figure \ref{fig:J0835-4510_results}, but for PSR J1257$-$1027.}
    \label{fig:J1257-1027_results}
\end{figure*}

\begin{figure*}
    \centering
    \includegraphics[width=\textwidth]{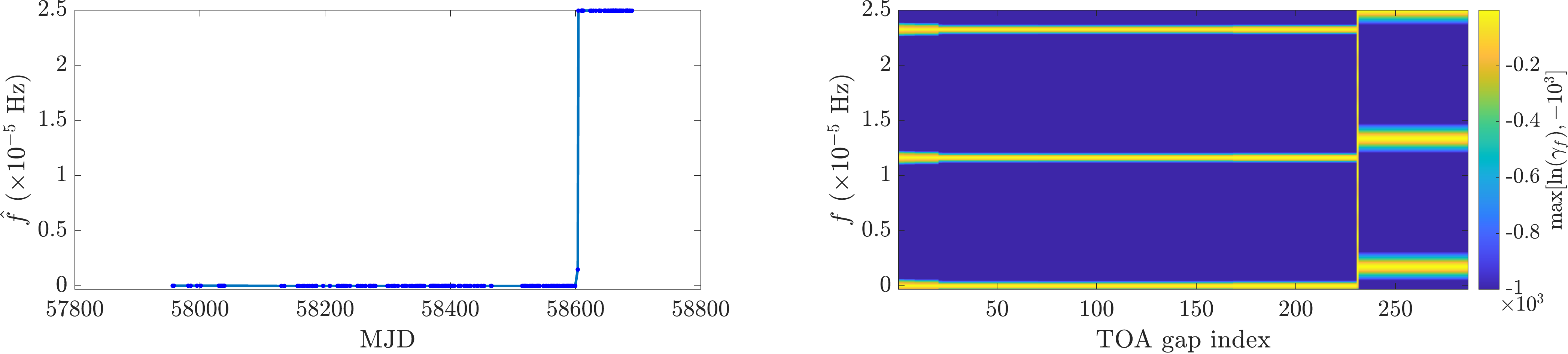}
    \caption{As in Figure \ref{fig:J0835-4510_results}, but for PSR J1452$-$6036.}
    \label{fig:J1452-6036_results}
\end{figure*}

\begin{figure*}
    \centering
    \includegraphics[width=\textwidth]{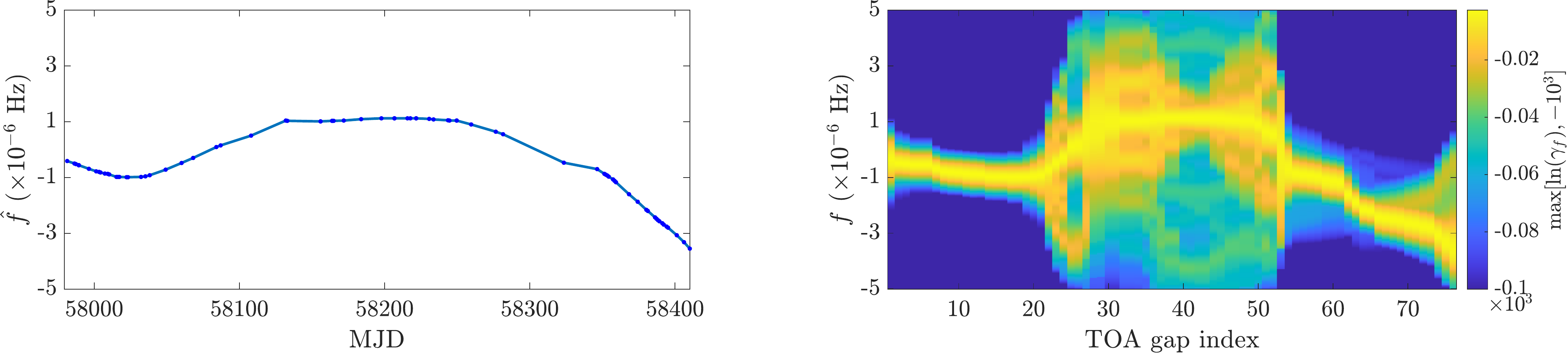}
    \caption{As in Figure \ref{fig:J0835-4510_results}, but for PSR J1622$-$4950.}
    \label{fig:J1622-4950_results}
\end{figure*}

\begin{figure*}
    \centering
    \includegraphics[width=\textwidth]{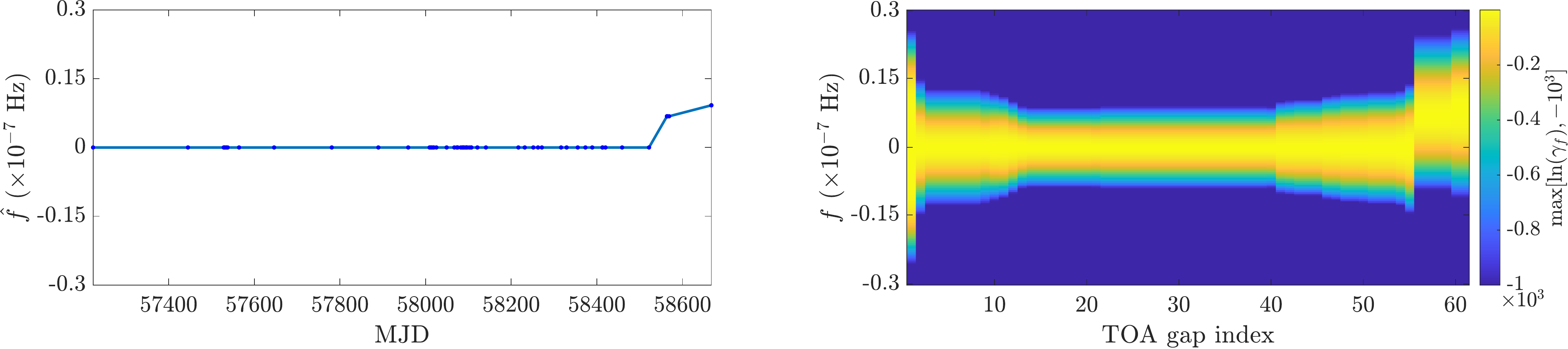}
    \caption{As in Figure \ref{fig:J0835-4510_results}, but for PSR J1703$-$4851.}
    \label{fig:J1703-4851_results}
\end{figure*}

\begin{figure*}
    \centering
    \includegraphics[width=\textwidth]{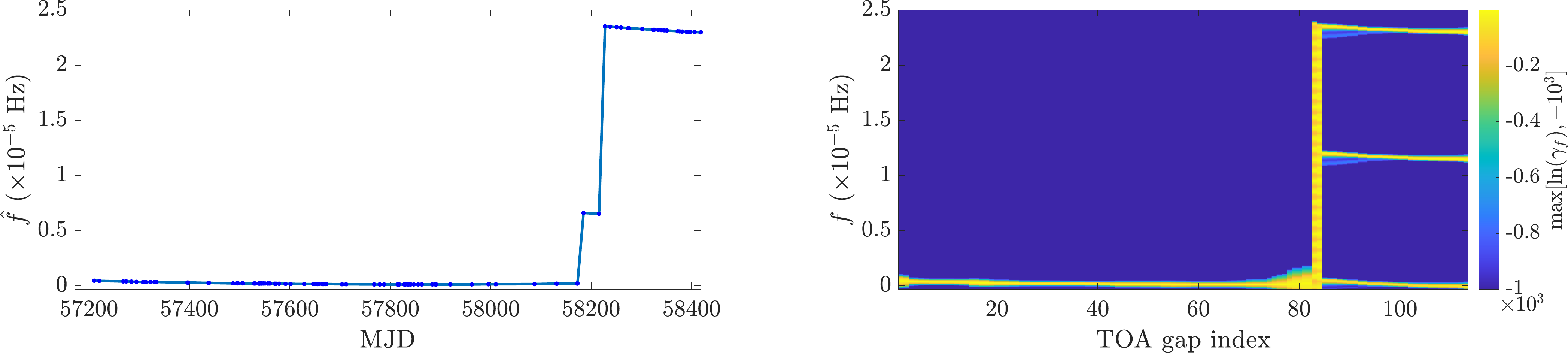}
    \caption{As in Figure \ref{fig:J0835-4510_results}, but for PSR J1709$-$4429.}
    \label{fig:J1709-4429_results}
\end{figure*}

\begin{figure*}
    \centering
    \includegraphics[width=\textwidth]{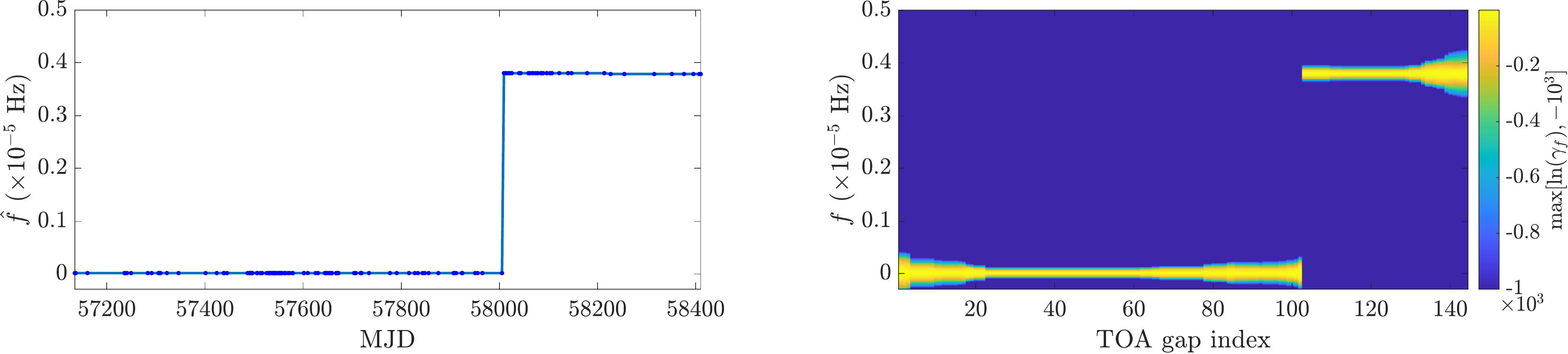}
    \caption{As in Figure \ref{fig:J0835-4510_results}, but for PSR J1731$-$4744.}
    \label{fig:J1731-4744_results}
\end{figure*}

\begin{figure*}
    \centering
    \includegraphics[width=\textwidth]{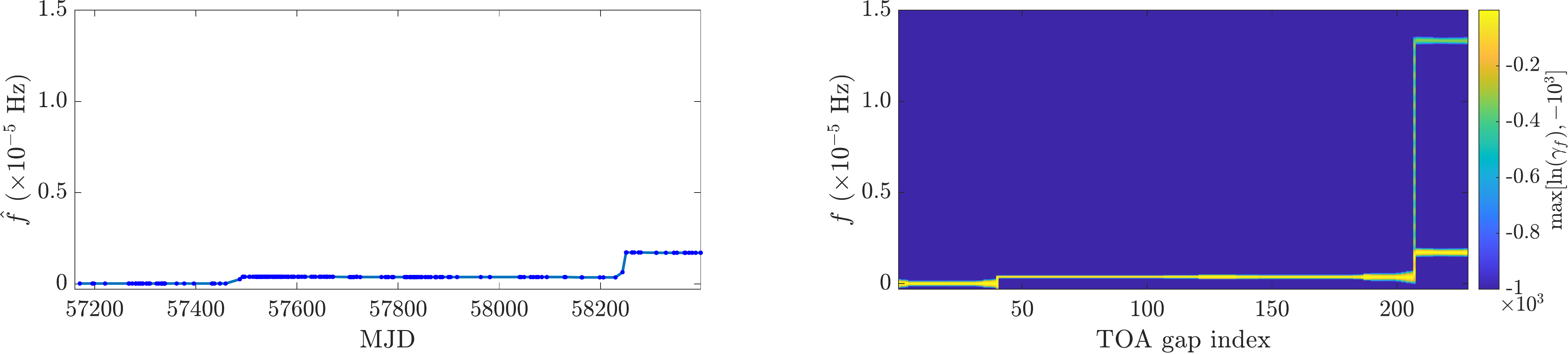}
    \caption{As in Figure \ref{fig:J0835-4510_results}, but for PSR J1740$-$3015.}
    \label{fig:J1740-3015_results}
\end{figure*}

This appendix collects Figs. \ref{fig:J0835-4510_results}--\ref{fig:J1740-3015_results} showing the sequence of most likely frequencies $\hat{f}(t_n)$ and heatmaps of the frequency posterior distributions $\gamma_f(t_n)$ for the eight pulsars which are followed up with glitch parameter estimation analyses as described in Section \ref{sec:results}.
The structure of each figure is essentially the same: the left panel shows $\hat{f}(t_n)$ as a function of the MJD, and the right panel shows $\ln[\gamma_f(t_n)]$ as a function of ToA gap index.
The values of $\ln[\gamma_f(t_n)]$ have been clipped below to aid readability.
The vertical axis in both cases extends over the full $f$ range in the DOI for each analysis.
In some cases $\gamma_f(t_n)$ displays multiple peaks; see Section \ref{subsec:results_J0835} and \citet{dunnEffectsPeriodicityObservation2021} for further discussion of this phenomenon.
Both $\hat{f}(t_n)$ and $\gamma_f(t_n)$ are obtained using the forward-backward algorithm \citep{rabinerTutorialHiddenMarkov1989}.
We also remind the reader that $\hat{f}(t_n)$ is the sequence of most likely states at each timestep (i.e. it is constructed from the sequence of modes of the posterior distribution of states) -- it is not the most likely sequence of states, which may instead be calculated using the Viterbi algorithm \citep{rabinerTutorialHiddenMarkov1989}.
However, the difference between these two sequences is typically small \citep{melatosPulsarGlitchDetection2020}, and so we prefer to use $\hat{f}(t_n)$.

\begin{figure*}
    \centering
    \includegraphics[width=0.5\textwidth]{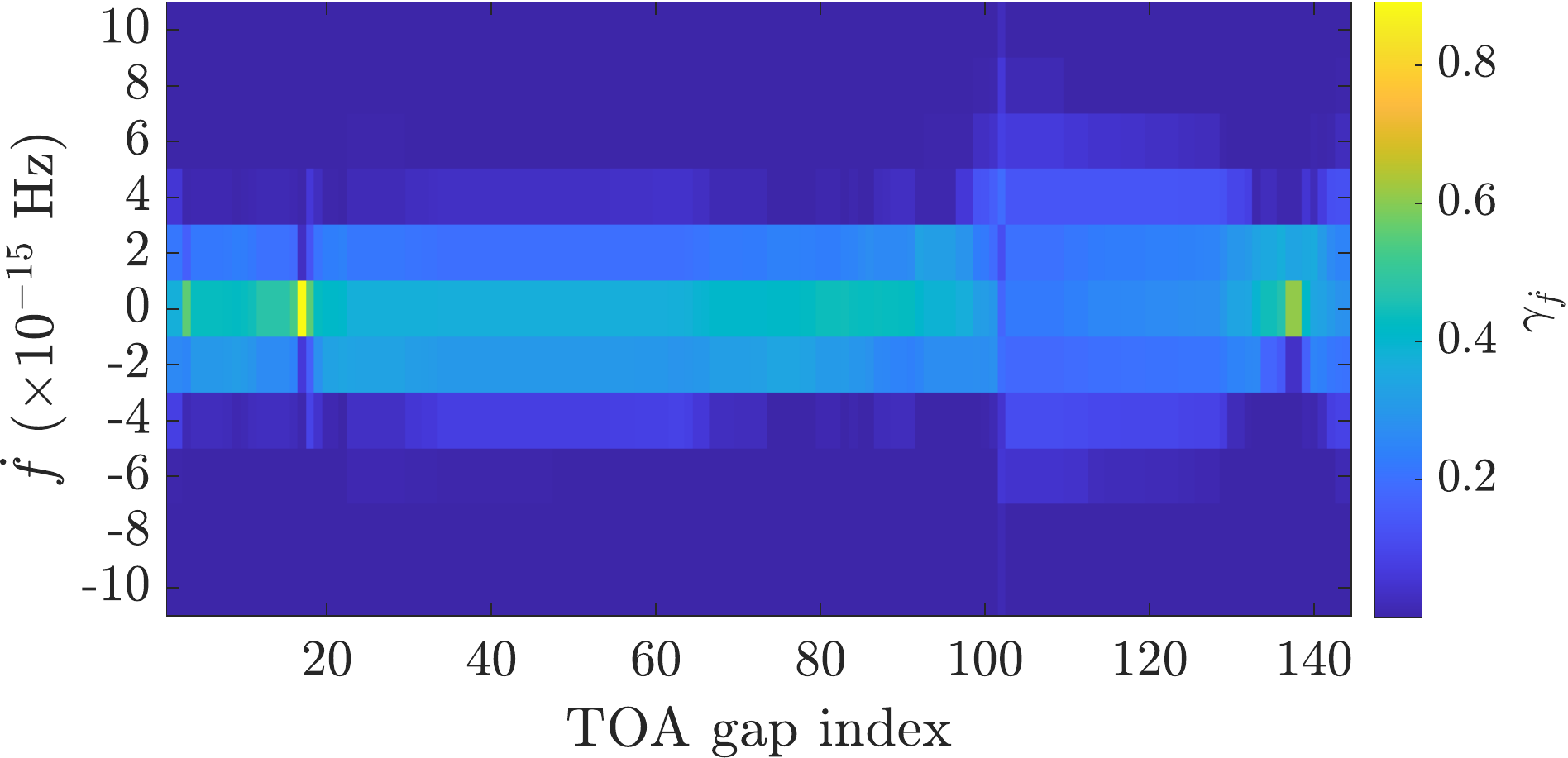}
    \caption{Heatmap of posterior frequency derivative probabilty $\gamma_{\dot{f}}(t_n)$ for the HMM follow-up analysis of J1731$-$4744.}
    \label{fig:J1731-4744_fdot_posterior}
\end{figure*}

We also include in Fig. \ref{fig:J1731-4744_fdot_posterior} an illustrative plot showing the frequency derivative posterior distribution $\gamma_{\dot{f}}(t_n)$ from the follow-up analysis of PSR J1371$-$3744. 
As mentioned in Section \ref{sec:results}, we do not include equivalent plots for every pulsar, as the coarse discretisation of $\dot{f}$ in the DOI leads to relatively uninformative $\dot{f}$ posteriors.
In the exemplar plot, $\gamma_{\dot{f}}(t_n)$ shows support over a significant fraction of the DOI, particularly after the glitch occurs at the 103rd ToA, making it difficult to make useful inferences about the evolution of $\dot{f}$ over the dataset.
Note that the heatmap shows $\gamma_{\dot{f}}$, not its natural logarithm, unlike Figs. \ref{fig:J0835-4510_results}--\ref{fig:J1740-3015_results}.

%%%%%%%%%%%%%%%%%%%%%%%%%%%%%%%%%%%%%%%%%%%%%%%%%%

% Don't change these lines
\bsp	% typesetting comment
\label{lastpage}
\end{document}